\documentclass[12pt,preprint]{aastex}
\shorttitle{Broadband GRB Behavior During $\gamma$-ray Emission}
\shortauthors{Yost et al.}

\def\lsim{\mathrel{\rlap{\lower4pt\hbox{\hskip1pt$\sim$}}
    \raise1pt\hbox{$<$}}}                % less than or approx. symbol
\def\gsim{\mathrel{\rlap{\lower4pt\hbox{\hskip1pt$\sim$}}
    \raise1pt\hbox{$>$}}}                % greater than or approx. symbol
\begin{document}

\title{Exploring Broadband GRB Behavior During $\gamma$-ray Emission}

\author{
Yost,~S.~A.\altaffilmark{1},
Swan,~H.~F.\altaffilmark{1},
Rykoff,~E.~S.\altaffilmark{1}, 
Aharonian,~F.\altaffilmark{2},
Akerlof,~C.~W.\altaffilmark{1},
Alday,~A.\altaffilmark{3},
Ashley,~M.~C.~B.\altaffilmark{4}, 
Barthelmy, S.\altaffilmark{5},
Burrows,~D.\altaffilmark{6},
Depoy,~D.~L.\altaffilmark{7},
Dufour,~R.~J.\altaffilmark{8},
Eastman,~J.~D.\altaffilmark{7},
Forgey,~R.~D.\altaffilmark{9},
Gehrels, N.\altaffilmark{5},
G\"o\u{g}\"u\c{s},~E.\altaffilmark{10},
G\"{u}ver, T.\altaffilmark{11},
Halpern,~J.~P.\altaffilmark{12},
Hardin,~L.~C.\altaffilmark{9},
Horns,~D.\altaffilmark{2},
K{\i}z{\i}lo\v{g}lu,~\"{U}.\altaffilmark{13},
Krimm, H. A.\altaffilmark{5,14},
Lepine,~S.\altaffilmark{15},
Liang,~E.~P.\altaffilmark{8},
Marshall,~J.~L.\altaffilmark{7},
McKay,~T.~A.\altaffilmark{1},
Mineo,~T.\altaffilmark{16},
Mirabal,~N.\altaffilmark{12},
\"{O}zel,~M.\altaffilmark{17},
Phillips,~A.\altaffilmark{4}, 
Prieto,~J.~L.\altaffilmark{7},
Quimby,~R.~M.\altaffilmark{18},
Romano,~P. \altaffilmark{19},
Rowell, G.\altaffilmark{2},
Rujopakarn,~W.\altaffilmark{1},
Schaefer,~B.~E.\altaffilmark{20}, 
Silverman,~J.~M.\altaffilmark{21},
Siverd,~R.\altaffilmark{7},
Skinner,~M.\altaffilmark{5},
Smith,~D.~A.\altaffilmark{1,22},
Smith,~I.~A.\altaffilmark{8},
Tonnesen,~S.\altaffilmark{12},
Troja,~E.\altaffilmark{16},
Vestrand,~W.~T.\altaffilmark{21},
Wheeler,~J.~C.\altaffilmark{18},
Wren,~J.\altaffilmark{23},
Yuan,~F.\altaffilmark{1},
Zhang,~B.\altaffilmark{24}
}

\altaffiltext{1}{University of Michigan, 2477 Randall Laboratory, 450
        Church St., Ann Arbor, MI, 48104, sayost@umich.edu, hswan@umich.edu,
        erykoff@umich.edu, akerlof@umich.edu,
        tamckay@umich.edu, wiphu@umich.edu, donaldas@umich.edu, yuanfang@umich.edu}
\altaffiltext{2}{Max-Planck-Institut f\"{u}r Kernphysik, Saupfercheckweg 1,
        69117 Heidelberg, Germany, Felix.Aharonian@mpi-hd.mpg.de,
        horns@mpi-hd.mpg.de, rowell@mpi-hd.mpg.de}
\altaffiltext{3}{Boeing LTS, AMOS Observatory, Kihei, Maui, Hawaii,  andrew.alday-iii@boeing.com, mark.a.skinner@boeing.com}
\altaffiltext{4}{School of Physics, Department of Astrophysics and Optics,
        University of New South Wales, Sydney, NSW 2052, Australia,
        mcba@phys.unsw.edu.au, a.phillips@unsw.edu.au}
\altaffiltext{5}{NASA/Goddard Space Flight Center, Greenbelt, MD 20771, scott@lheamail.gsfc.nasa.gov, gehrels@gsfc.nasa.gov, krimm@milkyway.gsfc.nasa.gov}
\altaffiltext{6}{Department of Astronomy, Penn State, 525 Davey Lab,
University Park, PA 16802, dburrows@astro.psu.edu}
\altaffiltext{7}{Department of Astronomy, Ohio State University, 140 West 18th Avenue, Columbus, OH, 43210, depoy@astronomy.ohio-state.edu, jdeast@astronomy.ohio-state.edu, marshall@astronomy.ohio-state.edu, prieto@astronomy.ohio-state.edu, siverd@astronomy.ohio-state.edu}
\altaffiltext{8}{Rice University, Department of Physics and Astronomy
6100 South Main, MS 108, Houston, TX, 77251-1892, rjd@rice.edu, liang@spacibm.rice.edu, ian@spacibm.rice.edu}
\altaffiltext{9}{Hardin Optical Company, P.O. Box 219, Bandon, OR 97411, bob@hardinoptical.com, larry@hardinoptical.com}
\altaffiltext{10}{Sabanc{\i} University, Orhanl{\i}-Tuzla 34956 Istanbul, Turkey, ersing@sabanciuniv.edu}
\altaffiltext{11}{Istanbul University Science Faculty, Department of Astronomy
        and Space Sciences, 34119, University-Istanbul, Turkey, 
        tolga@istanbul.edu.tr}
\altaffiltext{12}{Columbia University, Columbia Astrophysics Lab, 550 W. 120th St. Mail Code 5230, New York, NY 10027-6601, jules@astro.columbia.edu, mirabal@astro.columbia.edu, stonnes@astro.columbia.edu}
\altaffiltext{13}{Middle East Technical University, 06531 Ankara, Turkey,
        umk@astroa.physics.metu.edu.tr}
\altaffiltext{14}{Universities Space Research Association, 10211 Wincopin Circle, Suite 500, Columbia, MD  21044-3432, krimm@milkyway.gsfc.nasa.gov}
\altaffiltext{15}{Department of Astrophysics, Division of Physical Sciences, American Museum of Natural History, Central Park West at 79th Street, New York, NY 10024 lepine@amnh.org}
\altaffiltext{16}{INAF, Via U. La Malfa 153, 90146 Palermo, Italy
teresa.mineo@ifc.inaf.it, nora@ifc.inaf.it}
\altaffiltext{17}{\c{C}anakkale Onsekiz Mart \"{U}niversitesi, Terzio\v{g}lu
        17020, \c{C}anakkale, Turkey, m.e.ozel@comu.edu.tr}
\altaffiltext{18}{Department of Astronomy, University of Texas, Austin, TX
        78712, quimby@astro.as.utexas.edu, wheel@astro.as.utexas.edu}
\altaffiltext{19}{Osservatorio Astronomico di Brera,
Via Bianchi 46, 23807 Merate LC, Italy, patrizia.romano@brera.inaf.it}
\altaffiltext{20}{Department of Physics and Astronomy, Louisiana State
        University, Baton Rouge, LA 70803, schaefer@lsu.edu}
\altaffiltext{21}{Astronomy Department
, University of California at Berkeley, Berkeley, CA 94720-3411, JSilverman@astro.berkeley.edu}
\altaffiltext{22}{Guilford College, 5800 West Friendly Ave., Greensboro, NC 27410, dsmith4@guilford.edu}
\altaffiltext{23}{Los Alamos National Laboratory, NIS-2 MS D436, Los Alamos, NM 87545, vestrand@lanl.gov, jwren@nis.lanl.gov}
\altaffiltext{24}{Physics Department, University of Nevada, Las Vegas, NV 89154, bzhang@physics.unlv.edu}

\begin{abstract}
The robotic ROTSE-III telescope network detected prompt optical
emission contemporaneous with the $\gamma$-ray emission of {\em Swift}
events GRB\,051109A and GRB\,051111. Both datasets have continuous
coverage at high signal-to-noise levels from the prompt phase onwards,
thus the early observations are readily compared to the {\em Swift}
XRT and BAT high energy detections. In both cases, the optical
afterglow is established, declining steadily during the prompt
emission. For GRB\,051111, there is evidence of an excess optical
component during the prompt emission. The component is consistent with
the flux spectrally extrapolated from the $\gamma$-rays, using the
$\gamma$-ray spectral index. A compilation
of spectral information from previous prompt detections shows that
such a component is unusual. The existence of two prompt optical
components --- one connected to the high-energy emission, the other to
separate afterglow flux, as indicated in GRB\,051111 --- is not
compatible with a simple ``external-external'' shock model for the GRB and
its afterglow.

\end{abstract}
\keywords{gamma rays:bursts}

\section{Introduction}

Gamma-ray burst (GRB) early emission observations have become routine
since the launch of the {\it Swift} satellite \citep{gcgmn04}.  This
satellite has provided prompt triggers to events since early 2005,
whereby ``prompt'' designates ``during $\gamma$-ray emission''. With
the combination of such triggers and the increasing number of automated
rapid-response telescopes, the GRB field now has several examples of
optical lightcurves that begin during, or within seconds after, the
$\gamma$-ray emission.

Broadband prompt emission is one of the least-understood aspects of
GRB phenomena. The first prompt optical detection, GRB\,990123
\citep{abbbb99}, exhibited an optical flare that was interpreted as
the signature of a reverse shock passing through the relativistic
ejecta \citep[however, for another interpretation, see][]{lcbs99}. Reverse
shock emission was expected to be common \citep{sp99a}, so it has come
as a surprise that nearly all rapidly-detected GRB afterglows show
scant evidence of it. There are alternatives to the standard
``internal shocks'' formulation for prompt emission. Such models
include, e.g., external shocks \citep{mr93} or magnetic reconnection
\citep{mrp94,thompson94,usov94} as the mechanism to release energy as
$\gamma$-rays. The nature of GRB prompt emission is best investigated
in conjunction with prompt observations at lower frequencies, with
ongoing measurements at the same frequency to connect to the
longer-lasting, and better understood, afterglow.

The {\it Swift} XRT's early X-ray observations have revealed a nearly
standard morphology seen in most bursts' X-ray afterglow
\citep{nkgpg06,owogpv06}. The typical early X-ray afterglow includes
two surprises: a stage of relatively slow decay preceding the faster
decline, known from pre-{\em Swift} observations, hours to days
post-burst, and flaring well after the cessation of $\gamma$-ray
emission. To connect this interesting early behavior to the later
afterglow evolution, it is essential to compare such high-energy
emission to lower-energy evolution. Such comparisons elucidate which
features also occur at low energies, indicating a process affecting
the entire early afterglow rather than a separate high-energy
component \citep[see GRB\,050801,][]{rmysa06}.

There is a small but growing sample of events for which it is possible
to compare the very early optical lightcurve with X-ray emission (or
in the case of prompt optical detections, the GRB emission itself). To
date no consistent connection between prompt optical observations and the
contemporaneous $\gamma$-rays has emerged \citep[e.g., see the
discussion in][]{rykaa05}. We present here two new cases of
contemporaneous optical and high-energy observations. For GRB\,051109A,
the initial optical detection is prompt with respect to the
$\gamma$-rays, and is followed by significant overlap with X-ray
observations. GRB\,051111 does not have early X-ray observations, but
the optical lightcurve has significant temporal overlap (several
detections) with the {\it Swift} BAT $\gamma$-ray detections.

The structure of the paper is as follows. The events and our
observations are described in
\S\,\ref{sec:observations}. \S\,\ref{sec:reductions} gives technical
details for the data reduction, and
\S\S\,\ref{results:main},\ref{results:opt2fl}, and \ref{results:hi2fl}
detail the optical and high-energy transformations to spectral flux
densities. \S\S\,\ref{results:optfeatures} and
\ref{results:hifeatures} indicate the key features of the optical and
high-energy lightcurves, respectively. We discuss the lightcurves in
the context of the fireball model of
afterglows. \S\,\ref{firemodfeatures} summarizes important spectral
and temporal predictions of this model. The subsequent sections
discuss the lightcurves and broadband comparisons.
\S\,\ref{1109a_XO_main} compares optical and X-ray in GRB\,051109A. It
is divided into several subsections: \S\,\ref{1109a_X_1stto2nd}
examines the relative complexity of the X-ray during the first hour,
as compared to the steadily declining optical,
\S\,\ref{discuss:1109a_8hbreak} discusses the data near 0.5 day,
suggestive of achromatic steepening, while
\S\,\ref{discuss:1109abreak} looks at explanations and
\S\,\ref{discuss:1111latebreak} notes the similarity of the
GRB\,051111 optical lightcurve break near 1 ksec.
\S\,\ref{1111prompt} analyzes the prompt optical emission of
GRB\,051111, which includes a flux excess while high-energy emission
is detected. This excess is compared to other prompt
cases. \S\,\ref{sec:conclusions} summarizes the conclusions.

We use $\alpha$ to indicate temporal decay indices, and $\beta$ for
spectral indices, with flux density $f_{\nu} \propto t^{\alpha}\nu^{\beta}$. To
designate a spectral region, subscripts ``OPT'', ``X'', and
``$\gamma$'' indicate an index for the optical, X-ray, and
$\gamma$-ray bands respectively. A spectral index spanning two regions
is indicated with both, e.g., $\beta_{OPT-X}$ for the spectral
index interpolating between the optical and X-ray frequencies.

In the following, X-ray fluxes are measured in the band from 0.2--10
keV, and $\gamma$-ray fluxes correspond to 15--150 keV (in the
observed frame). The spectral shapes of higher frequency bands are
reported as photon indices, $\Gamma$ ($dn/d\nu \propto
\nu^{\Gamma}$). Note that the spectral index $\beta$ is related to the
photon index $\Gamma$, by $\beta = 1 - \Gamma$.

\section{Observations}\label{sec:observations}

The optical observations presented were taken by three
observatories. The instruments are described, followed by a
description of the instruments' responses to the events, and the XRT
response to GRB\,051109A.

The ROTSE-III array is a worldwide network of 0.45~m robotic,
automated telescopes, built for fast ($\sim 6$ s) responses to GRB
triggers from satellites such as HETE-2 and {\em Swift}.  They have a
wide ($1.\arcdeg85 \times 1.\arcdeg85$) field of view (FOV) imaged onto Marconi
$2048\times2048$ back-illuminated thinned CCDs, and operate without
filters, with sensitivity from approximately 400 to 900 nm. ROTSE-IIIb
is located at McDonald Observatory in Texas. The ROTSE-III systems are
described in detail in \citet{akmrs03}.

The MDM Observatory is located at Kitt Peak, Arizona. It includes the
1.3m and the 2.4m Hiltner telescope. The 2.4m has three CCDs, with
FOVs from 3.3 to 9.6 arcmin, which were used for the GRB\,051109A
observations. RETROCAM is the Retractable Optical Camera for
Monitoring, an Apogee ALTA E55 ($1152 \times 770$ pixels, with a scale
of $0\farcs26$ per pixel). Wilbur is a LORAL front side-illuminated
CCD ($2048 \times 2048$ pixels, with a scale of $0\farcs17$ per
pixel). Echelle is an SITe thinned, back side-illuminated CCD ($2048
\times 2048$ pixels, with a scale of $0\farcs28$ per pixel). These
CCDs operate with standard filters. The 1.3m has an SITe backside
illuminated CCD (17$'$FOV, with $0\farcs508$ per pixel), with a Harris
$R$ filter used for the late observation of GRB\,051111. Further
details for all instruments are available at the MDM
website\footnote{{\tt http://mdm.kpno.noao.edu/}}

The RUCCD instrument is installed on the 3.67 m Advanced
Electro-Optical System (AEOS) telescope at the Air Force Maui Optical
and Supercomputing (AMOS) site, located at 10,033 ft in Haleakala,
Hawaii. The camera is a $2048 \times 2048$ front-illuminated CCD,
covering a 45$''$ FOV. The instrument incorporates an
extensive set of optics options (polarizers, gratings, and filters),
which includes standard $V$, $R$ and $I$ filters. The RUCCD system is described in detail in \citet{sdlsh05}.

\subsection{GRB\,051109A}\label{obs1109}

On 2005 November 9, {\em Swift} detected GRB\,051109A ({\em Swift}
trigger 163136) at 01:12:20 UT. The position was distributed as a
Gamma-ray burst Coordinates Network (GCN) notice at 01:12:49 UT, with
a $3\arcmin$ radius error box, 29~s after the start of the burst
\citep{tbbcg05}. The burst had a duration of $36\pm2$~s (90\% duration,
15--350 keV), with a fluence of
$2.1\times10^{-6}\,\mathrm{erg}\,\mathrm{cm}^{-2}$ in the 15--150 keV
band \citep{fabbc05}. \citet{qfhrw05} measured an absorption redshift
of 2.346 for the event with the HET telescope, a few hours after the
burst.

ROTSE-IIIb responded automatically to the GCN notice, with the
first exposure starting at 01:12:52.7 UT, 32~s after the burst onset and
before the cessation of $\gamma$-ray activity.  The automated
scheduler began a program of ten 5-s exposures, ten 20-s exposures,
and 202 60-s exposures.  Near real-time analysis
of the ROTSE-III images detected a $15^{th}$ magnitude fading source
at $\alpha=22^h01^m15\fs3$, $\delta=+40\arcdeg49\arcmin23\farcs3$
(J2000.0) that was not visible on the Digitized Sky
Survey\footnote{{\tt http://archive.stsci.edu/cgi-bin/dss\_form}} red
plates. This was reported via the GCN Circular e-mail exploder within
9 minutes of the burst~\citep{rsryq05}.

{\it Swift} slewed immediately to the burst position, and the XRT
began X-ray observations 120~sec after this trigger
\citep{tbbcg05}. An uncatalogued X-ray source was detected at
$\alpha=21^{h}01^{m}15\fs24$, $\delta=+40\arcdeg49\arcmin23\farcs2$,
(J2000.0) with an estimated uncertainty of 3.5$''$ (90\% confidence
level), 0.7$''$ from the ROTSE coordinates.  This position takes into
account the correction for the misalignment between the telescope and
the satellite optical axis \citep{mpcca06}.  Due to orbital pointing
constraints, no XRT observations were made from
$t\approx200$--$3000$~sec, following which the field was visited
continually for 16 days, for a total of $\sim$283 ksec in 18
observations.

The MDM Observatory began $r$-band observations 38 minutes after the
burst, following the initial ROTSE GCN report. 23 exposures were taken
of the GRB field, spanning a total of 2.5 hours. Over the next 12 days
there were 4 further followup observations. 

MDM reobserved the afterglow of GRB\,051109A on 2006 June 29.456 UT
(mid-exposure), approximately 20 Msec after the event. The
observations consisted of four 600s R-band exposures acquired under
superb ($0^{\prime\prime}\!.8$) seeing conditions. At the location of
the afterglow, we detect a faint, extended object that we interpret to
be the host galaxy of this event.

\subsection{GRB\,051111}\label{obs1111}

On 2005 November 11, {\em Swift} detected GRB\,051111 ({\em Swift}
trigger 163438) at 05:59:41 UT. The position was distributed as a
Gamma-ray burst Coordinates Network (GCN) notice at 06:00:02 UT, with
a $3\arcmin$ radius error box, 20.5~s after the start of the
burst. The Moon's pointing constraint prevented {\it Swift}'s
narrow-field instruments from being brought to bear on the GRB
position immediately, and there are no early X-ray data
\citep{sbigh05}. The burst had a 90\% duration of $47\pm1$~s (15--350
keV), but there is extended emission to $>$80~s, and the burst's
fluence was $3.9\times10^{-6}\,\mathrm{erg}\,\mathrm{cm}^{-2}$
(15--150 keV) \citep{kabbc05}. \citet{hpfsr05} measured an absorption
redshift of 1.55 for the event with the HIRES instrument at the Keck
telescope, an hour after the burst.

ROTSE-IIIb responded automatically to the GCN notice in 6.4~s with the
first exposure starting at 06:00:08.4 UT, 26.9~s after the burst onset and
before the cessation of $\gamma$-ray activity.  The automated
scheduler began a program of 10 5-s exposures, 10 20-s exposures,
and 272 60-s exposures.  Near real-time analysis
of the ROTSE-III images detected a $13^{th}$ magnitude fading source
at $\alpha=23^h12^m33\fs2$, $\delta=+18\arcdeg22\arcmin29\farcs1$
(J2000.0) that was not visible on the Digitized Sky
Survey\footnotemark[2] red
plates, which we reported via the GCN Circular e-mail exploder within
8.3 minutes of the burst~\citep{rsrs05}.

The RUCCD instrument responded to the GCN notice, with its first
observation at 06:31:27 UT, 32 minutes after the burst. A series of
30-s observations of the GRB and a standard star were performed until
07:08:03 UT, using the $VRI$ filters, and initially reported in \citet{ss05}.

The MDM 1.3m telescope performed observations 1 day after the GRB
event. 75 minutes of $R$-band exposures were obtained over a period of
2 hours.

\section{Data Reductions}\label{sec:reductions}

\subsection{Optical Data Reductions}

ROTSE-III images were reduced and processed using the RPHOT pipeline,
with routines based upon DAOPHOT \citep{stetson87}. Objects were
identified via SExtractor \citep{ba96} and calibrated astrometrically
and photometrically with the USNOB1.0 catalog. The method is fully
described in \citet{qryaa05}, for the case of a well-separated
counterpart, such as GRB\,051111. The final result is a set of
PSF-fit photometric data.

The ROTSE-III instruments have 3$''$.25 pixels, and in the case of
GRB\,051109A the optical transient (OT) is partially blended with a
nearby ($8''$) 17th magnitude star at $\alpha=22^h01^m14\fs60$,
$\delta=+40\arcdeg49\arcmin25\farcs3$. It was therefore necessary to
remove this contaminating source prior to measuring the flux of the
OT, especially in images where the OT's flux has faded to or below a
similar level. To accomplish this, we first constructed a deep
reference image using ROTSE-III data obtained well after the OT had
faded below our detection limits. We measured the position and
brightness of the contaminating star on this frame. The star is then
removed from a given image by subtracting the locally determined PSF
scaled to the appropriate flux level at the star's position.

With the contaminating source removed, the GRB\,051109A OT light curve
was extracted as described in \citet{qryaa05}. The PSF-fit and larger
aperture light curves flatten out after $\sim$\,1 ksec post-burst.
However, the light curve resulting from the smallest aperture
($1\sigma$ radius, $\approx 3.5''$), which does not significantly overlap the
contaminating star, instead continues fading at the same rate.  This
demonstrates that residual light from the nearby star remains despite
our efforts to remove it. As the behavior of the $1\sigma$ radius
aperture should have little contamination from the neighboring star,
and since its light curve agrees with the behavior of the MDM data
obtained simultaneously (see below), we adopt these results as the
best estimate of the GRB\,051109A OT light curve. There are no
significant quantitative differences between the $1\sigma$ radius and
PSF-fit flux estimates in frames when the OT was brighter than 17th
mag. The estimated additional flux error due to slight misplacements
of these small apertures is negligible and is not included in the
results.

We have no data on afterglow color information for either GRB\,051109A
or GRB\,051111 at early times. Thus, no additional color corrections
for $R$-band equivalence have been applied to the ROTSE-III unfiltered
data, and the magnitudes quoted are then treated as $R$-band and
referred to as ``$C_{R}$''.

The MDM data were processed using standard IRAF/DAOPHOT procedures.
Aperture photometry was performed in a $1\farcs5$ radius aperture
(average seeing was a $\sim$\,$1\farcs4$ FWHM) centered on the OT and
nearby field stars. All GRB\,051109A observations, except the final
two, were performed with Sloan or Gunn $r$ filters.  With no $r$
standards in the MDM field, the GRB\,051109A $r$ observations are
converted to an $R$-equivalent value and presented in Table
\ref{tab:photometry} as ``$r_R$''. This is accomplished using
differential photometry with respect to two USNOB1.0 $R$ standards in
the field (at $\alpha$, $\delta$ $=22^h01^m15\fs663$,
$+40\arcdeg48\arcmin19\farcs01$ and $22^h01^m10\fs444$,
$+40\arcdeg49\arcmin50\farcs16$). The single GRB\,051111 $R$-band
point was calibrated with 5 USNOB1.0 stars within 3$'$ of the OT.

The late (host) observation of GRB\,051109A used the same aperture
size as the early OT observations ($1\farcs5$). As the seeing was good
($0\farcs9$), and the galaxy appears compact, the aperture includes
the total light contribution of the galaxy. The measurement
corresponds to $R_{host}$ = $23.70\pm0.16$, but does not change within
the uncertainties when using aperture sizes from
$0\farcs8$--$2\farcs1$. Assuming the scaling of star-formation rate
(SFR) to UV continuum luminosity of \citet{k98}, the implied
uncorrected SFR in the host galaxy is $\approx 18$ M$_{\odot}$
yr$^{-1}$. This value indicates a moderate SFR, when compared to
starburst galaxies, as in previous cases \citep[e.g.,][]{chg04}.
Using an astrometric solution derived from a set of suitable field
stars also present in earlier images, we find that the afterglow
position is offset from the center of this galaxy by less than
$0^{\prime\prime}\!.11$ ($\pm 0^{\prime\prime}\!.15$).

The RUCCD data were reduced using IRAF procedures and processed for
aperture photometry with the IRAF 2.12.2 {\tt qphot} package. A
standard star, BD+42 4211, from the Bright Northern BVRI
Standards\footnote{{\tt
http://stupendous.cis.rit.edu/tass/refs/skiff\_photom.html}} was
observed between the two sets of GRB observations (the first set from
2--2.5 ksec, and the second from 3--4 ksec post-trigger). Its
observations were used to reference the $VRI$ GRB observations to
absolute photometry.

The final optical magnitudes from these instruments are listed in the
tables. Table\,\ref{tab:photometry} gives the GRB\,051109A results, and
Table\,\ref{tab:photometry2} the GRB\,051111 results.

\subsection{High-Energy Data Reductions}

The GRB\,051109A XRT data were first processed by the {\it Swift} Data
Center at NASA/GSFC into Level 1 products (event lists). This includes
the initial 2.5 sec image, and the following Windowed Timing (WT) and
Photon Counting (PC) observations (until 3440 sec, and after 3440 sec,
post-trigger respectively) The event lists were further processed with
the XRTDAS (v1.7.1; in FTOOLS) to produce the final cleaned
event lists. In particular, the {\tt xrtpipeline} (v0.9.9) applied
calibration and standard filtering and screening criteria. Temporal
intervals during which the CCD temperature was higher than $-47$
$^\circ$C were cut out, and hot and flickering pixels were removed.
An on-board event threshold of $\sim$0.2 keV was also applied to the
central pixel, to reduce most of the background due to either the
bright Earth limb or the CCD dark current.  The events selected for
analysis had XRT grades 0--12 and 0--2 for PC and WT data,
respectively \citep[for the {\it Swift} XRT nomenclature,
see][]{bhnkw05}.

The GRB\,051109A WT data were extracted in a rectangular region 40
pixels long along the image strip and 20 pixels wide. The afterglow
was sufficiently intense to cause pile-up in the PC mode data until
the third orbit.  To account for this effect the source events were
extracted in an annulus with a 20-pixel outer radius ($\sim47$\arcsec)
and a 3-pixel inner radius.  These radii were derived by comparing the
observed and nominal PSF.  For the PC data collected after the third
orbit, the entire circular region (20-pixel radius) was used, instead.
The selected extraction regions correspond to $\sim 93.5$ \% (WT),
$\sim 47.4$ \% (piled-up PC), and $\sim 88.5$ \% (non piled-up PC) of
the XRT PSF. These fractions were applied to correct the measurements
to the full count rate. The background was measured from data within
an annulus (radii 70 and 130 pixels) centered on the source (PC mode),
and within a rectangular box (40$\times$20 pixels) far from background
sources (WT mode).

Spectra of the source and background were extracted in the regions
described in above from the first orbit event files.  Ancillary
response files were generated with the task {\tt xrtmkarf} within
FTOOLS, with RMF (v007) spectral redistribution matrices.  The 0.5--10
keV WT data and 0.2--10 keV PC data were simultaneously fit to an
absorbed power law model, with Galactic $N_{\rm H}^{\rm Gal}=1.75
\times 10^{21}$ cm$^{-2}$, and a free $N_{\rm H}$ parameter at the GRB
redshift. A free constant factor was introduced to take into account
the decrease of the mean flux between WT and PC data.  The result is a
photon index $\Gamma=2.06\pm0.09$ and a column density of $N_{\rm
H}=7.91_{-3.68}^{+4.24} \times 10^{21}$ cm$^{-2}$ (90\% errors for one
interesting parameter). There is no evidence for strong spectral
changes after the first orbit. The GRB\,051109A count rates were then
converted to 0.2--10 keV unabsorbed fluxes (presented in Table
\ref{tab:XRT}) using the best spectral fit.

BAT data were used for $\gamma$-ray comparisons in {\it Swift}
bursts. For these, the event files from the public archives were
analyzed with the BATTOOLS and XSPEC11 software packages\footnote{{\tt
http://swift.gsfc.nasa.gov/docs/swift/analysis/}} . Using the
appropriate housekeeping files, mask weighting was determined with
{\tt batmaskwtevt}, then signal-to-noise binned lightcurves were made
with {\tt batbinevt}. Spectral response ({\tt .pha} and {\tt.rsp})
files were created for specific time intervals using {\tt batbinevt,
batupdatephakw, batphasyserr} and {\tt batdrmgen}. XSPEC11 fit photon
indices and returned unabsorbed flux values (15--150~keV) from the
{\tt .pha} and {\tt.rsp} files. The comparison of count rates and
returned fluxes yields a general conversion factor that can be applied
to transform a count rate lightcurve to fluxes.

\section{Results}\label{results:main}

For these two bursts, there is a significant overlap between the early
optical observations and much higher energy emission (X-ray or
$\gamma$-ray). Temporal and spectral comparisons can elucidate whether
the higher and lower energy emission come from the same spectral
component or emitters. GRB\,051109A has good temporal overlap between
the optical and X-ray observations. The GRB\,051111 early optical
lightcurve has significant overlap with the $\gamma$-ray lightcurve,
allowing a comparison of the lightcurve evolution in both bands. This
section describes the conversions required for these comparisons from
the photometry in Tables\,\ref{tab:photometry} and
\ref{tab:photometry2}, or {\em Swift} data.

We determine the onset of $\gamma$-ray emission, referred to as
$t_{GRB}$, by examining the BAT lightcurves for these events. The
adopted onset for GRB\,051109A is UT 01:12:15.5. This is compatible
with the time of initial BAT detections reported by
\citet{fabbc05}. The adopted onset time for GRB\,051111 is UT
05:59:39, congruent with the report of {\it Suzaku}'s initial
$\gamma$-ray detection time \citep{ysotf05}.

\subsection{Optical Transformations to Flux Density}\label{results:opt2fl}

In the following, all comparisons from the optical to the higher
frequencies use $R$ band optical flux densities corrected for Galactic
extinction unless explicitly indicated otherwise. All magnitudes from
Tables \ref{tab:photometry} and \ref{tab:photometry2} are converted
using the effective frequencies and zero point fluxes of
\citet{b79}. In particular, the $R$-equivalent values are converted as
if they were $R$. This includes the ROTSE $C_R$ and MDM $r_R$
magnitudes. These values are corrected for extinction with the
prescription of \citet{sfd98}. The resulting corrections are 0.511 and
0.433 mags of Galactic extinction in the $R$ band for GRB\,051109A and
GRB\,051111 respectively. These corrections are applied to all
$R$-equivalent ($C_R$ and $r_R$) observations.

\subsection{Summary of Key Optical Features}\label{results:optfeatures}

\begin{figure}
\epsscale{1.0}
\plotone{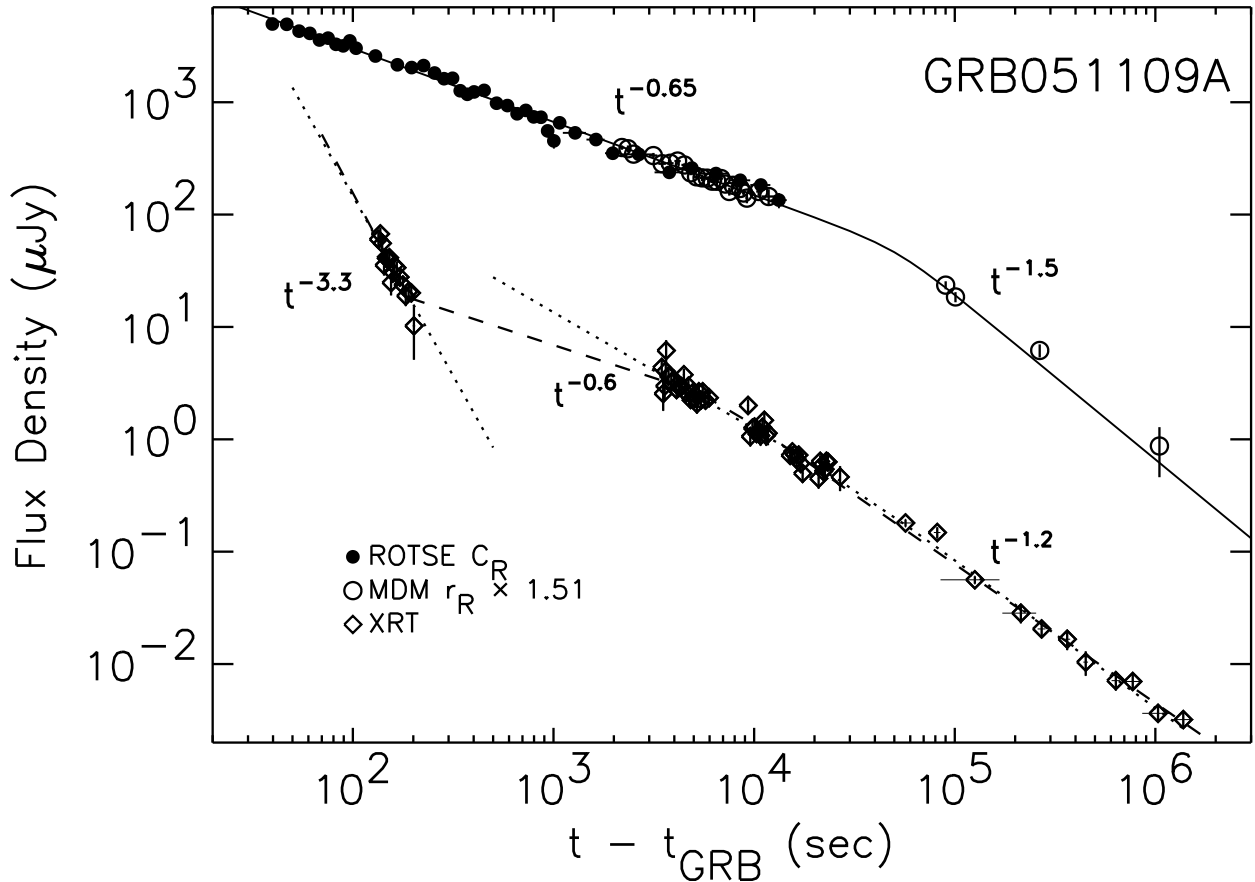}
\caption{GRB\,051109A optical and X-ray early lightcurves.  The ROTSE
magnitudes and MDM of Table\,\ref{tab:photometry} are converted to
flux densities (and corrected for by 0.511 mag of Galactic extinction,
see \S\,\ref{results:opt2fl}), and the XRT flux density conversions
are described in \S\,\ref{results:hi2fl}. The adopted onset time
$t_{GRB}$ is UT 01:12:15.5. The optical lightcurve is fit by a double
power law (see Table\,\ref{tab:fits}) from $t^{-0.65}$ to $t^{-1.5}$
with a break time of $50\pm10$ ksec, plus a constant host term. The
fit includes a color term between the MDM $R$-equivalent $r$ and the
ROTSE $R$-equivalent unfiltered values, thus the MDM points are
multiplied by the fitted factor of 1.51 on the plot. The fitted host
value of $1.4\pm0.2$ $\mu$Jy is subtracted from the optical lightcurve
(and the final point is not plotted), in order to show the evolution
of the optical afterglow light. The first XRT orbit shows a steep
decay discontinuous with subsequent X-ray evolution. The later XRT
data can be fit by a double power law (dotted lines showing the
unlinked fits of data before and after the orbital gap), or by a
triple power law (dashed line, where the shallow segment of $t^{-0.6}$
has no data to anchor it). The latter approach shows the average flux
evolution through the data gap. The X-ray and the optical data, taken
together, show an steepening consistent with achromaticity around 0.5
days (see \S\,\ref{discuss:1109a_8hbreak}). Post-break, the decays
and spectral index can be explained by an ISM or windlike model with
the cooling break above the X-ray (\S\,\ref{1109a_XOspec}). Pre-break,
the temporal decays are too shallow to be easily explained by the
fireball model, although long-duration smooth energy injection is a
possibility, see \S\,\ref{discuss:1109abreak}.}\label{fig:fig1109a}
\end{figure}

Figure \ref{fig:fig1109a} gives the GRB\,051109A optical
lightcurve. The host flux is significant by 1 Msec. Its value,
$1.4\pm0.2$ $\mu$Jy, is fit from the entire lightcurve and subtracted
in the figure to show the afterglow evolution.

As the dataset combines $C_R$ and $r_R$ data, we allow for a color
offset between them, which is evident during the significant overlap
from $\sim$ 2--20 ksec post-onset. The term is a simple multiplicative
factor ($1.5\times$, see Table \ref{tab:fits}) for the flux density,
fitted with the optical lightcurve. It is applied to the MDM data as
plotted in Figure 1.  A color term of $1.5\times$ (0.4 mag) is
substantial. However, it ited $r$ observations calibrated to $R$
standards to unfiltered data, effectively two different optical bands
near $R$.  The MDM data during the overlap hints at a shallower
evolution than the ROTSE data, but separate fits show this is not a
significant result. The figure shows that with the color term, the two
lightcurves are in agreement.

The first 5-second ROTSE data point for GRB\,051109A is
contemporaneous with $\gamma$-ray emission. The optical afterglow of
GRB\,051109A follows a power law decline at this time, behavior
previously observed in GRB\,050401 \citep{rykaa05}. In this case, the
early optical afterglow steepens between the early observations (until
10 ksec), and the next night. A double power law fit exhibits an
initial decline of $t^{-0.652 \pm 0.008}$ until a time $52 \pm
9$~ksec, when it transitions to $t^{-1.5 \pm 0.2}$ (see Table
\ref{tab:fits}).  This describes the data adequately, as seen in
Fig. \ref{fig:fig1109a}, but has $\chi^2 = 96$ for 62 degrees of
freedom (DOF). We note that the sparser data after 1 day leaves the
final decay less well-constrained. Without the final observation,
there would be no host constraint, and the lightcurve would appear to
have a more shallow decay and an earlier optical break.

\begin{figure}
\epsscale{1.0}
\plotone{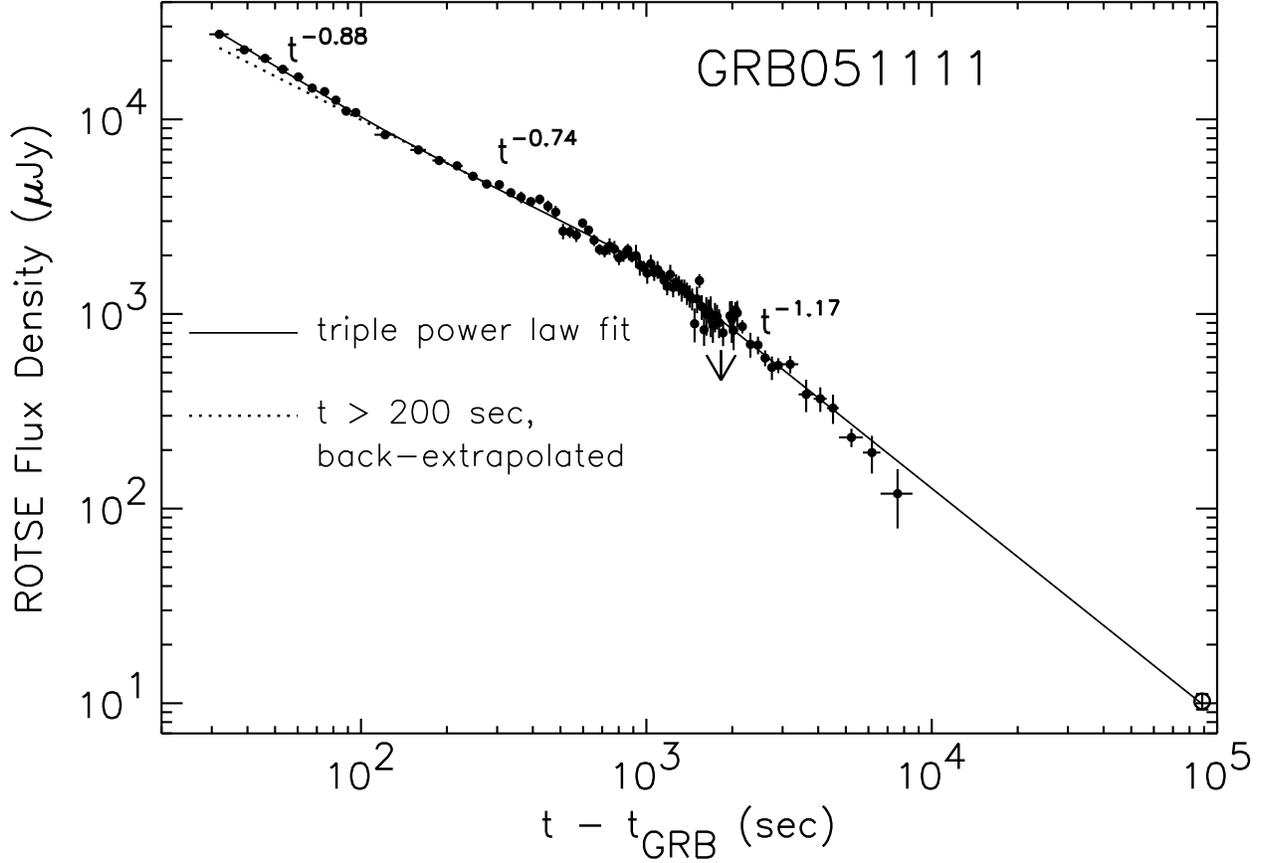}
\caption{GRB\,051111 optical ROTSE lightcurve. The ROTSE magnitudes of
Table \ref{tab:photometry2} are converted to flux densities (corrected
for 0.433 mag of Galactic extinction, see \S\,\ref{results:opt2fl}). A
ROTSE observation not detected at $> 3\sigma$ significance is given as
a $3\sigma$ upper limit and indicated by the arrow. A single late MDM
observation is included as an unfilled point, with no color offset
applied. The adopted onset time $t_{GRB}$ is UT 05:59:39. The optical
is fit with a triple power law, as described in
\S\,\ref{results:opt2fl} and reported in Table\,\ref{tab:fits} (solid
line). The optical decay from 100 to 1000 sec post-onset is
$t^{-0.74\pm0.01}$. The break after 1000~sec is by $\Delta\alpha =
-0.43\pm0.03$, which does not fit any expected spectral or jet break
in the simple fireball model. It may indicate a similar process as
that which produces the shallow break in GRB\,051109A
(Fig. \ref{fig:fig1109a}), see \S\,\ref{discuss:1111latebreak}. During
the prompt $\gamma$-ray emission, lasting until $\sim$80--100 sec
post-onset (see Fig. \ref{fig:1111_gam}), the optical light decays
more rapidly than after its end. The {\it dashed} line shows the
back-extrapolation of the lightcurve's fitted power law evolution
after $\sim$ 150 sec.}\label{fig:1111_opt}
\end{figure}

Figure \ref{fig:1111_opt} shows the GRB\,051111 optical lightcurve. The
optical is well-described by power law evolution, with breaks. There
are two breaks, from an initial decay to a slightly flatter evolution
at about 2 minutes post-onset, and then to a steeper decay near
0.5~hr post-onset. The latter break is discussed in
\S\,\ref{discuss:1111latebreak}, while the former resembles an optical
excess during the $\gamma$-ray emission, discussed in
\S\,\ref{1111prompt}.

Using least-squares fitting, the optical lightcurve is characterized
by a triple power law (Table \ref{tab:fits}). Initially, it decays as
$t^{-0.88\pm0.02}$, then at $t=120\pm20$~sec, it flattens slightly to
$t^{-0.74\pm0.01}$, and finally at $1100\pm90$~sec, it steepens to
$t^{-1.17\pm0.02}$. This fit includes the final MDM $R$-band point
with no color term relative to the ROTSE $C_{R}$, but the results are
not affected if the fit is performed without the MDM
point. The overall decay does not change between 10 and 100 ksec
post-burst \citep[although there may be some fluctuation about the
power law, see][]{blphu06}.

The existence of the initial steeper decay and the first optical
break is visible to the eye and statistically well-established. The
triple power law fit gives a $\chi^2$ of 78 for 83 DOF. A double power
law can also be fit to the data, resulting in a single break at
1500~sec and a $\chi^2$ of 98 for 85 DOF. Both fits yield acceptable
$\chi^2$, but the double power law fit has larger systematic
residuals. An $F$-test indicates that the decrease in $\chi^2$ by
allowing the early, third power law segment is statistically
significant at a confidence level greater than 0.999.

In both Figures \ref{fig:fig1109a} and \ref{fig:1111_opt}, the optical
decay has begun a steady decline by the first observations. These two
events are well-sampled examples where the optical decay is
established in a form like the fireball model afterglow during the end
of the prompt emission, only tens of seconds after the start of the
GRB. This is in contrast to some other cases, where the optical is
rising during \citep[GRB\,050820A,][]{vwwag06} or after
\citep[GRB\,060605,][]{srsq06} the prompt emission.

\subsection{High-Energy Data Transformations to Flux Densities}\label{results:hi2fl}

We use the GRB\,051109A XRT fluxes from Table \ref{tab:XRT}.
These de-absorbed X-ray fluxes are converted to flux densities using the
spectral fit. The spectrum's power law photon index ($2.06\pm0.09$)
yields a weighted mean frequency of 5.7$\times10^{17}$~Hz and a
conversion factor of $4.35\times10^{4}$Jy per erg~s$^{-1}$cm$^{-2}$
(adding a further uncertainty of 5\%) to convert to flux density.

We obtained the $\gamma$-ray data for these events from the BAT
observations in the {\em Swift} public archive. These are reduced with
the standard suite of BATTOOLS (see \S\,\ref{sec:reductions}) to produce
spectral fits, count rates, and fluxes as needed.

For GRB\,051109A, the BAT fluxes were analyzed by checking spectral fits
over the entire burst ($-10$ to 50~sec post-onset), yielding $\Gamma
= 1.50\pm0.15$. Analysis from $-10$ to 5 and $5$ to 50~sec showed
consistent photon indices, so there is no evidence for spectral
evolution throughout this burst. Thus spectral parameters fitted from
the entire burst were used to produce fluxes and flux densities in the
BAT band from count rates. This method was also used in analyzing
other events, as described in \S\,\ref{1111context}.

For GRB\,051111, the BATTOOLS spectral fits show that
the BAT spectral index softens from $1.22\pm0.04$ ($-5$ to 10~sec) to
$1.48\pm0.07$ (10 to 100~sec). The times of interest all occur during
the later interval, and this interval's value of of $\Gamma$ is used
to convert from count rate to flux density. 

BAT and XRT fluxes were also compared (for GRB\,051109A) using the
BAT flux spectrally extrapolated to the XRT energy range. The best
determination of the BAT spectrum is obtained via XSPEC11 fits to the
entire data set. The BAT fit parameters allow a flux extrapolation in
XSPEC11 from the total BAT count rate to the flux in the XRT energy
band. The extrapolation can be applied to each BAT data point in the
count rate light curve.

\subsection{Summary of Key High Frequency Features}\label{results:hifeatures}

Figure \ref{fig:fig1109a} shows the GRB\,051109A XRT lightcurve for
this event. There is a data gap from 205 to 3400~sec, due to orbital
constraints. We examined the data both by fitting before and after the
gap independently, as well as fitting the entire dataset. All the
results are presented in Table \ref{tab:fits}.

The first orbit data, from 134.0 to 204.5~sec post-onset, shows a
steep decline as $t^{-3.2\pm0.4}$ while during this stage the optical
decays at a much slower rate. The later observations, from 3.4 to
1400~ksec, show the X-ray flux declining more shallowly.

Considered in isolation, the X-ray data after the first orbit shows
similar behavior to the optical lightcurve, a double power law, but
with a shallower break. Initially decaying $\sim$\,$t^{-1}$, at
approximately 30~ksec, it would steepen to $t^{-1.3}$.  The
back-extrapolation of this fit is brighter than the X-ray flux level
at the end of the first orbit.  This requires a near-``plateau'' as
the overall flux evolution throughout the data gap (an unseen fourth
power law segment in the lightcurve).

The entire X-ray dataset can be fit by a triple power law: $t^{-3.3}$
shallowing to $t^{-0.6}$ at 200 sec, then steepening to $t^{-1.2}$ at
7 ksec. Due to data uncertainty, both fits, linked and unlinked across
the data gap, are statistically good ($\chi^2$ reported in
Table\,\ref{tab:fits}). The triple power law fit is driven by the data
gap: the data at 6~ksec does not show an evident break, and fitting
the data before and after the gap separately yields a fit at late
times that best matches ($t^{-1.32}$) just the late time behavior. The
fit residuals for the model over the entire XRT dataset appear to show
trends near the 6~ksec break and the late-time decay, however they are
not statistically unreasonable using the ``runs test'' for the signs
of residuals.  The interpretation of the data can accomodate
significant variation in the break time (30 vs. 7 ksec) and final
decay ($t^{-1.3}$ vs. $t^{-1.2}$).

We note that the early XRT decay differs from the behavior through the
data gap. GRB\,051109A XRT observations began after the end of
$\gamma$-ray emission. The prompt high-energy emission was spectrally
extrapolated to the XRT band for comparison.  Initally comparable in
flux, the XRT decay is steeper than that implied from the flux
extrapolations from the BAT during the period from $\sim$1 sec to 1
minute from the GRB onset. This would be consistent with the
interpretation of steep X-ray emission from high-latitude photons, at
the end of the GRB.

\begin{figure}
\epsscale{1.0}
\plotone{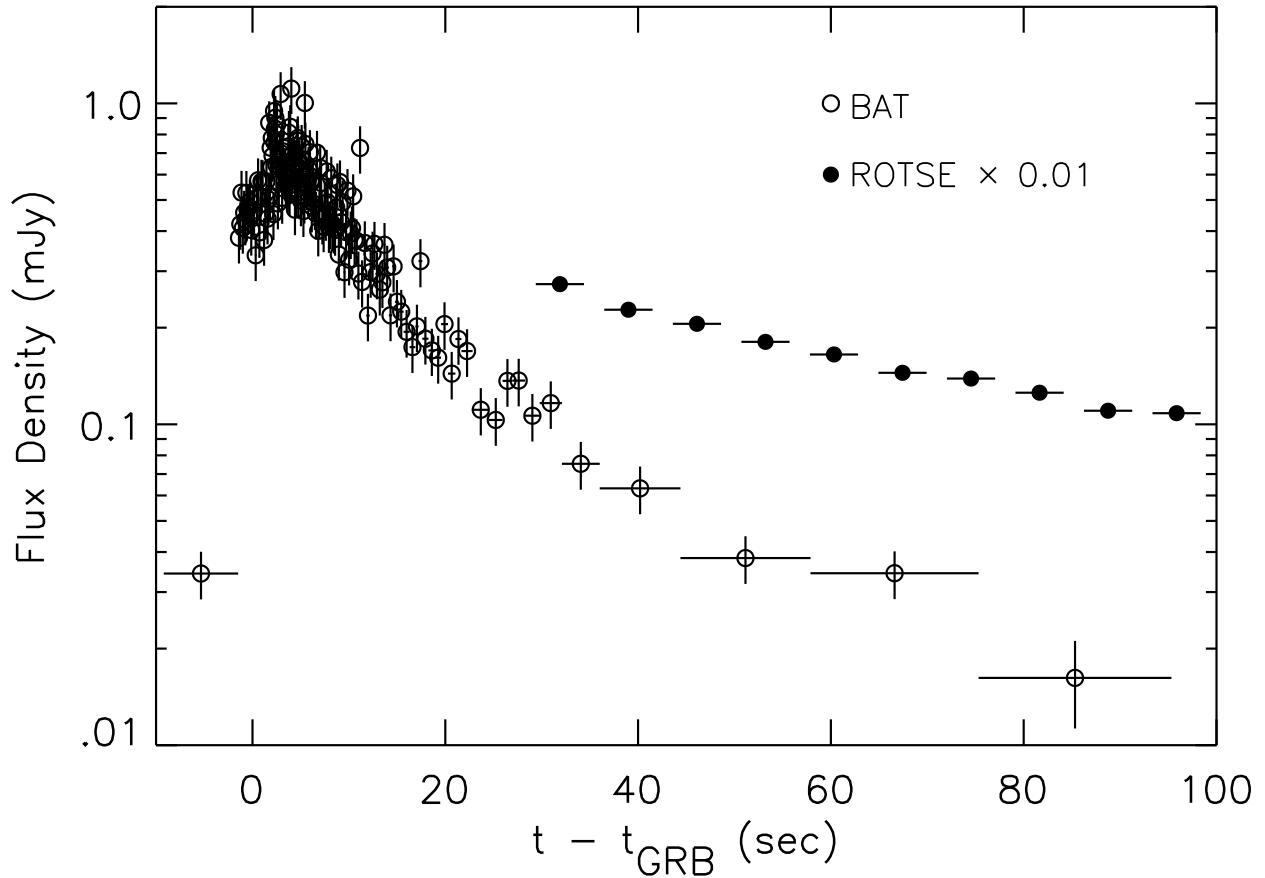}
\caption{GRB\,051111 {\it Swift}~BAT 15--150 keV $\gamma$-ray
lightcurve, compared to ROTSE-III prompt optical detections. The ROTSE
flux densities are as described in Fig. \ref{fig:1111_opt}, now scaled
by a factor of 0.01 for comparison, and the BAT flux density
conversions are described in \S\,\ref{results:hi2fl}.  The adopted
onset time $t_{GRB}$ is UT 05:59:39. The linear timescale allows the
point before $t_{GRB}$ to be shown, and thus that the onset matches
the beginning of $\gamma$-ray emission.  There is $\gamma$-ray
emission to approximately 80--100 sec post-onset. The prompt optical
flux declines more slowly than the smooth tail of $\gamma$-ray
emission.  }\label{fig:1111_gam}
\end{figure}

Figure \ref{fig:1111_gam} shows the GRB\,051111 BAT lightcurve. The
BAT lightcurve from 15--200~sec post-onset (a time range chosen for
comparison with the optical lightcurve which begins at 31.9~sec) is
fit as $t^{-1.50\pm0.07}$ ($\chi^2=41$ for 32 DOF). The temporal
behavior is fit using the count rate lightcurve, to avoid the
additional uncertainty in the conversion from count rate to flux
density. The count rate fit is normalized at 31.9~sec, with an
amplitude that corresponds to a flux density of $81.7\pm4.1$~$\mu$Jy
at a weighted mean frequency of $1.7\times 10^{19}$Hz.

\section{Fireball Model Features for Interpretation}\label{firemodfeatures}

Both GRB\,051111 and GRB\,051109A exhibit optical lightcurve breaks at
early times. To interpret these events, we use the simple fireball
model \citep{mr97, spn98}. This section describes the model's
predicted spectral and lightcurve evolution rate characteristics that
are relevant for optical and higher-energy frequencies \citep[][fully
reviews the model]{p05}. Later sections will demonstrate that the
optical and X-ray lightcurve breaks cannot be explained by the model
predictions outlined below.

The fireball describes the emission from a population of accelerated
electrons with Lorentz factor distribution $N(\gamma_e) \propto
\gamma_e^{-p}$. The afterglow spectrum has spectral breaks:
principally $\nu_m$, due to the mimimum Lorentz factor $\gamma_e$, and
$\nu_c$, the cooling frequency. There is also a self-absorption
frequency, but it is far below the optical when the optical transient
flux is decaying.

The electron index $p$ determines the synchrotron flux spectral
shape. Each spectral region has a spectral index $\beta$, with flux
density $f_{\nu}\propto\nu^{\beta}$. The index $\beta = (1-p)/2$ for
$\nu_m < \nu < \nu_c$, and $\beta = -p/2$ for the case when $\nu >
\nu_c$ and $\nu > \nu_m$. (When $\nu_c < \nu_m$, the spectral
shape is $\nu^{-1/2}$ for frequencies between them.) The frequencies
evolve in time depending upon geometry and the circumburst density
distribution $n(r)$. The standard assumptions for the density are $n$
constant (``ISM'' case), or $n \propto r^{-2}$ (``windlike'' case).

Each spectral region (relative to the break frequencies) and density
regime has a relation between the lightcurve temporal evolution and
$p$, e.g., $f_{\nu} \propto t^{3(1-p)/4}$ for $\nu_m < \nu < \nu_c$ in
the ISM case for spherical geometry. These relations require $p >
2$. Otherwise either the total energy diverges or there is a
high-energy cutoff that drives the flux evolution at a different rate.

The fireball model predicts lightcurve breaks due to both spectral
evolution from the cooling adiabatic shock, and hydrodynamic
transitions. While the fireball will produce rising lightcurves at low
frequencies (attenuating self-absorption, or the approach of the
spectral peak), above the peak, the flux decays. The high-frequency
lightcurves are expected to steepen.

Spectral evolution will produce chromatic breaks, with break
frequencies evolving typically $\sim t^{-0.5}$--$t^{-1.5}$. The
steepening follows a pattern, and is predicted to be shallow.

The cooling frequency, $\nu_c$, is expected to produce a break. It
evolves as $t^{-1/2}$ for an ISM density, and $t^{1/2}$ for a
windlike one. With an ISM density, the cooling frequency will start
high. As $\nu_c$ sweeps below a frequency, its lightcurve acquires a
dependence upon $\nu_c$ and steepens by $\Delta\alpha = -0.25$, so
higher frequencies will have steeper decays. A windlike density
follows the opposite pattern, as $\nu_c$ sweeps {\it up}, making
lightcurves shallower for frequencies $\nu > \nu_c$. Lightcurves of
frequencies $\nu_m < \nu < \nu_c$ are steeper by $\Delta\alpha =
-0.25$ than of frequencies where $\nu_m < \nu_c < \nu$.

The passage of $\nu_m$ steepens a decaying lightcurve when $\nu_c <
 \nu_m$. When $\nu_c < \nu < \nu_m$, the lightcurve decays shallowly
 (with $f_{\nu} \propto t^{-1/4}$) for both ISM and windlike cases. It
 then steepens to the decay rate above $\nu_c$ and $\nu_m$. This
 requires a very shallow initial decay. The passage of $\nu_c$ is the
 relevant case for changes in temporal evolution of lightcurves that
 decay faster than $t^{-1/4}$.

Hydrodynamic changes can also provide lightcurve breaks. As the whole
shock is affected, these breaks would be achromatic. The simple model
predicts a ``jet break'' due to observing the effects of collimated
ejecta. A (sharp-edged) cone of ejecta initially evolves
hydrodynamically as if it were isotropic. When the ejecta have slowed
sufficiently for the emission's beaming angle to be larger than the
ejecta's opening angle, there will be a break due to the ``missing
light'' compared to a spherical distribution of emitters. By geometric
arguments, this would require a steepening by $\Delta\alpha =
-0.75$. At approximately this time, the ejecta will begin expanding
significantly sideways, putting more energy into expansion and further
weakening the observed emission. The model expectation would be a
larger $\Delta\alpha$, leading to a final lightcurve decay of $t^{-p}$
(again, for $p > 2$) at all frequencies above the spectral peak.

In the simple formulation, a chromatic break would be shallow and due
to $\nu_c$, while an achromatic break would be strong and due to the
jet. As discussed below, the lightcurve breaks in the GRB\,051111 and
GRB\,051109A do not follow these predictions.

\section{Evidence for Long-Term Energy Injection in Optical and X-ray}\label{1109a_XO_main}

The GRB\,051109A optical and X-ray lightcurves overlap for several
decades in time. Although the data gaps allow some ambiguity, under
any interpretation, the comparison of optical and X-ray evolution
requires processes beyond the simple self-similar adiabatic fireball
model.

\subsection{The First Hour: X-ray and Optical Behavior Compared}\label{1109a_X_1stto2nd}

\S\ref{results:hifeatures} discusses the ambiguous measurements of the
 early X-ray lightcurve evolution. The data can be fit by the
 nearly-canonical triple power law shape, with the shallow segment
 during the orbital data gap. A fit to the data after the gap yields a
 later, shallower break (see Table\,\ref{tab:fits}).

As discussed in \S\ref{firemodfeatures}, the fireball model predicts a
shallow chromatic break or a strong achomatic break. However, the two
X-ray fits, when compared to the optical lightcurve, follow neither
pattern. 

First, if the X-ray afterglow follows the triple power law shape,
there are similar strong steepenings in the X-ray and optical bands
hours after the GRB, but the X-ray break occurs before the optical
one. This is not characteristic of a jet or any other hydrodynamic
transition.

The X-ray triple power law is suggested by the lightcurve
morphology from early analysis of XRT afterglows \citep[][with 27
events]{nkgpg06}. However, some XRT lightcurves do not follow that
pattern; in 40 cases \citet{owogpv06} identify several without the
shallow or ``hump'' phase. These include examples with a single power
law from early times onwards, either very steep (e.g., GRB\,050421),
or less so (e.g., GRB\,050721). There are also cases where flares
obscure where a ``shallow'' lightcurve segment may be (e.g.,
GRB\,050908). If GRB\,051109A's X-ray afterglow followed the
steep--shallow--steep pattern, the data gap would coincidentally
include the entire shallow segment. There is no information to
determine whether GRB\,051109A's X-ray decayed shallowly ($t^{-0.6}$)
from 200 to 3400 sec, or had a more complex lightcurve evolution.

The second possible interpretation treats the data before and after
the data gap (0.2--3 ksec) separately. Hours after the burst, it
yields a shallow break, at the same time as the shallow optical break
(thus not following the simple fireball predictions). The X-ray fit
after the gap back-extrapolates to a flux brighter than observed at $t
\approx 0.2$ ksec. This requires a more complex discontinuity in the
X-ray evolution. The optical has well-sampled steady decay during the X-ray's
data gap, which makes the implied discontinuous X-ray evolution difficult to explain in a broadband context.

Under this interpretation, during the data gap, the X-ray lightcurve
must brighten relative to its previous decay. This would be explained
if the X-ray afterglow rose between orbits. However, the afterglow
peak is already below the optical. The optical flux decay during the
XRT data gap shows that $\nu_m < \nu_{\mathrm{OPT}} < \nu_{X}$, and
the afterglow would already have risen at X-ray frequencies.

A flatter X-ray evolution before 3.4~ksec would be expected if $\nu_c$
dropped below the X-ray band at that time. A break passage coincident
with the end of a data gap would be surprising. This interpretation is
also unlikely due to the lack of evidence for change in X-ray spectrum
between the first orbit and the second; the spectral index steepens by
0.5 when $\nu_c$ passes.

One possibility is an unseen flare (at XRT frequencies, during the
data gap) that does not decline to the original underlying level
\citep[see rare examples in][Fig. 2]{owogpv06}. Such a flare would
have to have no effect upon the optical evolution, steady during this
time. This would be surprising as the similar X-ray and optical decays
appear to indicate a common emission source by the X-ray's second
orbit. Any effect that boosts the flux level in the X-ray would be
expected to have some effects in the optical. There is no physical
parameter in the fireball model upon which the flux at high
frequencies depends that the flux at lower frequencies (above the
spectral peak) does not. Specifically, if energy is injected to raise
the flux level, it will affect the entire spectrum.

By 40~sec post-onset, the optical remnant of GRB\,051109A has begun a
steady decline consistent with the synchrotron model from an external
shock (with the steady addition of energy). The X-ray emission is not
compatible with the afterglow until some time between 0.2 and
3.4~ksec. The establishment of emission from the assumed self-similar
forward shock appears more complex at high energies than at low
ones. 

\subsection{GRB\,051109A: X-ray and Optical Breaks Near 0.5 Day}\label{discuss:1109a_8hbreak}

The optical lightcurve for GRB\,051109A has a break approximately half
a day after the burst. The XRT lightcurve steepens on a
similar timescale. 

A lack of full coverage limits our knowledge of these breaks in both
bands. The optical lightcurve has sparse coverage after about 20 ksec
and has a strong host contribution by 1 Msec. This prevents a precise
measurement of the break time and post-break decay. The optical steepening
may be significant, but at $\Delta\alpha_{OPT}=-0.8\pm0.2$ it is also
consistent within 3\,$\sigma$ of a shallow ($\Delta\alpha=-0.25$)
cooling break. The XRT data gap (200--3400 sec) prevents a good
measurement of the segment before the break. The X-ray may transition
from a quite shallow ($t^{-0.6}$) decay around 2 hours, with
$\Delta\alpha_{X}=-0.64$, or it may have an initially steeper evolution
($t^{-1}$) with a break of $\Delta\alpha_{X}=-0.28$ near 9 hours. The
shape of the afterglow lightcurves is more uncertain than the
statistical error bars of model-fit decay rates and transition times
(Table\,\ref{tab:fits}).

There is a transition at both frequencies at a comparable time. If the
steepening were due to a break frequency passing from the X-ray to the
optical, the break would have to evolve at least as fast as
$\sim\nu\propto t^{-3.5}$ (if the X-ray break is around 7 ksec), or
even $\sim\nu\propto t^{-17}$ (if the X-ray break is at 34
ksec). This is faster than any break evolution expected in the
fireball model. With the uncertainty in break times, the optical and
X-ray lightcurves are consistent with an achromatic break time. Due to
the significant uncertainty in the optical steepening $\Delta\alpha$,
the lightcurves are consistent with having the same amount of
steepening $\Delta\alpha$.

Using the fits from Table\,\ref{tab:fits}, we find the spectral index
from the optical to the X-ray before and after the breaks near 0.5
days. The results are $\beta_{OPT-X}= -0.65\pm0.15$ ($-0.8\pm0.2$) at 6
(100) ksec. The consistency of $\beta_{OPT-X}$ across the observed
shallow breaks also points to the X-ray and optical breaks arising
from a single cause, producing an achromatic effect upon the spectrum
of a single emission source.

\subsubsection{Fireball Spectral Constraints}\label{1109a_XOspec}

The position of spectral breaks relative to the observed frequencies
can be constrained via $\beta$ and $\alpha$. Given the
uncertainties in the optical and X-ray decay rates and lightcurve
breaks, more than one type of fireball model could explain the
GRB\,051109A afterglow data.

After the breaks at $\sim$\,0.5 days, the
optical apparently decays more quickly than the X-ray, at
$t^{-1.5\pm0.2}$. However, there is significant uncertainty in this
decay rate; by refitting with various fixed decay indices $\alpha$, we
find the optical decay may be as shallow as $t^{-1.1}$ (3\,$\sigma$
significance). This would be consistent with the X-ray decay being
steeper than the optical decay by $\Delta\alpha\approx-0.25$ before
and after this break, if the X-ray was decaying as steeply as $t^{-1}$
at $t\approx3$ ksec, after the data gap.

As $\beta = -p/2 < -1$ when $\nu > \nu_m, \nu_c$, the relatively
shallow $\beta_{OPT-X}\approx -0.7$ points to $\nu_c > \nu_{OPT}$, with
$\nu_c$ above or just below the X-ray. After the 0.5 day breaks, if
the X-ray is decaying as $t^{-1.3}$ and the optical as $t^{-1.1}$,
$\nu_c$ would be between the optical and X-ray and the circumburst
density would be constant, like the ISM (in the windlike case, higher
$\nu$ do not decay more quickly than lower $\nu$). This would require
an electron energy index $p\approx2.4$, and point to the
interpretation that the X-ray was decaying more quickly than the
optical before the 0.5 day break as well. Alternatively, if both the
X-ray and optical are decaying $\sim t^{-1.3}$ after the break, the
shallow $\beta_{OPT-X}$ would indicate $\nu_m < \nu_{OPT} < \nu_{X} <
\nu_c$, which would be satisfied for a windlike medium with
$p\approx2.4$ and for an ISM medium with $p\approx2.7$.

\subsection{Does the Fireball Model Explain the Breaks Near 0.5 Day?}\label{discuss:1109abreak}

The GRB\,051109A afterglow lightcurves have shapes somewhere between
the two (X-ray) cases inferred from the data. The ordinary fireball
model does not easily explain either the case of simultaneous optical
and X-ray breaks (with a shallow $\Delta\alpha_{X}$) or that of both
$\Delta\alpha\approx -0.7$ (with the optical break later than the
X-ray).

If the breaks occur at the same time, they do not resemble the
expected achromatic jet break steepening. The X-ray break appears far
shallower (051109A XRT 2 in Table \ref{tab:fits} $\Delta\alpha \approx
-0.3$) than the minimum $\Delta\alpha=-0.75$ for a non-spreading jet
(\S\ref{firemodfeatures}). Moreover, the pre-break decays
(GRB\,051109A and XRT 2 in Table\,\ref{tab:fits}) are not well
explained by the fireball model. As the X-ray decay is apparently
steeper than the optical initially, the model indicated would be an
ISM density with $\nu_m < \nu_{OPT} < \nu_{c} < \nu_{X}$. Then the
quite shallow optical decay is difficult to interpret, as it indicates
an electron energy index $p=1.87\pm0.01$, a value $p<2$ resulting from
relations that assume $p>2$.

The alternate interpretation has an X-ray decay $\sim t^{-0.6}$ during
the data gap, and the X-ray break before the optical break. The
shallow early decay is still difficult to interpret. While the break
amplitudes $\Delta\alpha\approx-0.7$ (both bands) could be interpreted
as arising from a non-expanding jet, the break times are not
achromatic as expected. Interpreting the late X-ray decay as post-jet
also calls into question all afterglow interpretations. Decays of
$t^{-1.3}$ have been routinely observed in afterglows on $\sim$~day
timescales, and interpreted as spherical fireball behavior with
typical electron energy indices $p\approx2.4$.

Puzzling behavior has been observed in early broadband afterglow data.
Some cases show chromatic early lightcurve breaks, seen only in the
X-ray while the optical decay rate remains constant
\citep{pmbng06}. Others show simultaneous optical and X-ray breaks
\citep[e.g., GRB\,050525A and GRB\,050801,][]{bbbbc06, rmysa06}.  The
GRB\,050801 afterglow break was achromatic for optical and X-ray
frequencies, but did not have the characteristics of a jet break
\citep{rmysa06}. The nearest analogy to GRB\,051109A may be the
GRB\,050525A afterglow, which exhibited both optical and X--ray
steepening at $t\approx4$~hours, with break amplitudes differing
between the frequencies \citep{bbbbc06}. The final GRB\,050525A
lightcurve decays are quite steep (nearly $\sim t^{-2}$), and
\citet{bbbbc06} tentatively conclude it is a jet break. Such a steep
decay is not observed in GRB\,051109A.

\subsubsection{No Obvious Explanation by Complex $n(r)$}

Under either lightcurve interpretation, a more complex environment
cannot be easily used to produce the breaks. A change in density
gradient would be a hydrodynamic change, and would produce an
achromatic effect. The effect would be tiny for frequencies above both
$\nu_c$ and $\nu_m$. An achromatic break time appears to require an
X-ray decay initially steeper than optical, so $\nu_c$ would be
between the frequencies, and only a transition in the optical would be
expected. Disregarding the break times, it is quite difficult to
produce a transition from $t^{-0.65}$ to $t^{-1.2}$. If the latter is
an ISM model fireball observed between $\nu_m$ and $\nu_c$, it implies
$p\approx2.7$. Such a steep energy index would not produce a shallow
$t^{-0.65}$ decay even with a sharply rising density gradient
\citep[see][Table 5]{yhsf03}.

An alternative environmental effect such as variable extinction would
have to affect frequencies from optical to the X-ray in the same
fashion, which is not reasonable for known absorbers. Thus, environmental
changes do not provide a plausible solution.

\subsubsection{No Obvious Explanation by Changing Burst Parameters}

We checked whether changing the physical parameters of the fireball
could readily explain the observed lightcurves. Beyond the general
question of early energy injection, recent analyses of early chromatic
break cases have opened the question of changing microphysical
parameters \citep{pmbng06}.

As mentioned above (\S\,\ref{discuss:1109abreak}), a shallow
$t^{-0.65}$ decay could imply a hard $p<2$. A steepening of $p$ would
conceivably produce the lightcurve breaks (particularly if both bands
were indeed shallow early on, with the same transition to a final
decay rate). This possibility still presents a difficulty, as the spectral
index would also steepen. For a sufficiently significant steepening of
$p$, $\Delta\beta_{OPT-X}\sim-0.5$, and the ratio of X-ray to optical
flux would drop by more than a factor of 10 during the transition at
$t>20$\,ksec. This is not observed (Fig. \ref{fig:fig1109a}).

It is also difficult to produce the lightcurve breaks by changing the
electron or magnetic energy fractions, $\epsilon_e$ or
$\epsilon_B$. Even a shallow lightcurve break would require
significant parameter changes \citep[see the spectrum's dependences
upon physical parameters, summarized by][]{p05}. We note that generic
microphysical parameter changes have many degrees of freedom, if
$\epsilon_e$, $\epsilon_B$, and $p$ can vary independently. As such,
this is a poorly-constrained hypothesis for the various ``flattened''
early decays in GRB afterglows.

A more physically-motivated parameter change is the continuous
injection of energy into the forward shock. Such a change would in
general slow the lightcurve decay, causing a break when the injection
ends. This simple formulation would not be able to explain a
steepening in the X-ray before the optical, and the GRB\,051109A
afterglow lightcurves do not definitely have an achromatic
break. However, the similar timescales for the steepening of optical
and X-ray lightcurves suggest such a common cause.

Increased energy could be provided in a burst when the engine emits a
distribution of material with Lorentz factors distributed in a power
law \citep[$M(\gamma)\propto\gamma^{-S}$, e.g.,][]{sm00}. The gradual
``catching up'' of material to the slowing forward shock increases the
shock energy and decreases the flux decay rate. This model affects
both spectral regions above the peak (above and below $\nu_c$), giving
them a comparable amount of flattening. This explanation can provide
the appropriate level of pre-break flattening for GRB\,051109A, both
at optical and X-ray, with an influx of material with Lorentz factors
distributed as $M(\gamma)\sim\gamma^{-4}$.

\subsection{GRB\,051111 Optical Break at 0.5 hours: Similarity to GRB\,051109A}\label{discuss:1111latebreak}

The optical afterglow of GRB\,051111 shows a break at 1200~sec with
amplitude similar to the GRB\,051109A break (Table
\ref{tab:fits}). Its amplitude, $\Delta\alpha = -0.43 \pm 0.03$, is
too large to be due to the passage of the spectrum's cooling break. It
is also too shallow to be interpreted as a jet break. There is no
overlapping X-ray data for comparison, but it shares the
characteristics that make the GRB\,051109A afterglow break
incompatible with the simple fireball model.

The decay post-break is consistent with several simple fireball
scenarios. For GRB\,051111 at $t > 1200$~sec, the decay can be fit by an
ISM or windlike medium, with the optical above the peak and the
cooling frequency of the synchrotron spectrum. The decay is also
consistent with the optical above the spectral peak but below the
cooling frequency, although in the windlike case the decay rate would
imply a hard electron energy distribution, with $p\approx2$.

The RUCCD data (Table \ref{tab:photometry2}) give rough $VRI$
information at 1 hour post-burst. We used the data to constrain the
optical spectral slope at this time. The values were corrected by
0.537, 0.433, 0.314 magnitudes for Galactic extinction in $V$, $R$,
and $I$ respectively. The data did not have the precision required to
determine the spectral shape, but a fit to a power law for the three
points yields a $\beta_{OPT}$ of $-0.4 \pm 1.0$. This result is
consistent with all spectral scenarios for the density regimes
discussed above. It slightly favors the cases where the optical band
is still below the cooling frequency at this time, which give harder
optical spectra.

The decay pre-break is difficult to reconcile with the standard
fireball model expectations.  Using the $p > 2$ relations for this
shallow decay results in an inconsistent value of $p < 2$ for a windlike
medium in any spectral ordering, or for an ISM medium above the
cooling frequency.  The ISM medium model, with the optical between the
peak and $\nu_c$, would be just consistent with $p = 2$. As the ISM
model finds $p$ significantly above 2 post-break, it leaves the break
unexplained.

\subsection{Breaks From Halting Energy Injection?}

The afterglow of GRB\,051109A has a break observed at optical and
X-ray frequencies at a similar time, tens of ksec post-burst. Data
sampling gaps do not permit a definite determination of whether the
break was achromatic, yet even a near proximity in time suggests a
common cause. Geometric and environmental characteristics do not
provide a plausible explanation, but the end of a stage of smooth
energy injection could steepen lightcurves as observed.

The GRB\,051111 break may be due to similar causes as the GRB051109A afterglow
break. As in the GRB\,051109A case, the GRB\,051111 optical afterglow's
shallow decay also lasts substantially longer than the prompt
$\gamma$-ray detections (until 1200 sec cf. 80 sec). The GRB\,051111 break
can also be explained by continued energy injection from the GRB time
until the break time.

\section{GRB\,051111: A Prompt Optical Component Consistent with the extension of the $\gamma$-rays}\label{1111prompt}

We now discuss the high-energy comparison for the early GRB\,051111
optical observations. There is no early X-ray temporal overlap, but a
significant overlap with $\gamma$-ray observations. The ``90\% fluence
duration'' of the burst is T$_{90} = 47 \pm 1$~sec in the 15--350~kev
band \citep{kabbc05}, and the flux has a smooth decay from 10~sec
after the onset that extends to 80~sec post-trigger (see Figure
\ref{fig:1111_gam}).  From the onset $t_{GRB}$ (05:59:39, see
\S\ref{results:main}), three optical points are contained within
T$_{90}$, and seven are at $t - t_{GRB} < 80$~sec. We can compare not
just a single point to the $\gamma$-ray emission, but rather the
optical evolution to the $\gamma$-ray evolution, as fitted from their
lightcurves.

The optical lightcurve at $t < 125$~sec is brighter than the
extrapolation of the $\sim$125--1000 sec data to earlier times. The
decay before 125~sec is steeper than afterwards, which suggests an
``excess component'' during the GRB emission. The ``excess'' would be
the difference between the observed emission and the extrapolation of
the later, shallower emission to this early time. Taking the
difference between the data and the back-extrapolation of the power
law which dominates from $\approx 0.1$--$1$~ksec gives an estimate of
the excess component. The difference is well-fit by a power law decay
of $t^{-1.8\pm0.4}$.

The implied $t^{-1.8}$ decay rate of the excess could be expected for
a reverse shock component \citep{sp99a}, but it would be a surprising
coincidence for an excess to arise from a reverse shock component
lasting precisely the GRB timescale. The GRB decline is
$\alpha_{\gamma} = -1.50\pm0.07$ (\S\,\ref{results:hifeatures}), so it
is possible that the excess is correlated with the GRB emission
\citep[as in the case of GRB\,041219A,][]{vwwfs05}.  As the GRB has a
single peak with a smooth tail during the optical observations, it is
not possible to establish a correlation from lightcurve morphology.

As discussed in \S\,\ref{results:optfeatures}, the data permit a fit
by a double power law with a single break at $t\sim1$~ksec, which
would imply no early excess. The triple power law from which we infer
an excess is visible and statistically established by the fit
improvement. Yet due to the data's uncertainties, a fit to a double
power law plus an excess power law component will find an acceptable
fit with no early excess component. The properties of the excess
component cannot be constrained with such a general fit model.

As the above estimate of the early optical excess is consistent with
the $\gamma$-ray lightcurve decay, we attempt to refine the
component's estimation under the assumption that its decay is linked
to the contemporaneous $\gamma$-ray lightcurve. We consider the
optical flux density data for $t < 1$~ksec and the $\gamma$-ray count
rates from 15--150~sec. This dataset was fit to a function
$At^{\alpha_1} + Bt^{\alpha_2}$ (optical) and $Ct^{\alpha_1}$
($\gamma$-rays). A Monte Carlo method determined the fit parameter
uncertainties. Artificial datasets were generated, using the measured
values and uncertainties to form Gaussian distributions for each data
point, and then fit, yielding the distributions of the function's
parameters.

With 2000 trials, $\alpha_1 = -1.44\pm0.07$ and
$\alpha_2=-0.70\pm0.03$.  These are consistent with
the previous measurement of this phase's $\gamma$-ray decay and
$\alpha_2$ in the triple power law fit (Table\,\ref{tab:fits})
respectively. The estimated optical excess at 31.9~sec ($A$) is
$8.2\pm2.1$~mJy. The distribution for $A$ is nearly Gaussian, and
$A>0$ is significant at above the 3~$\sigma$ level. This estimation of
the optical excess agrees with the ``implied excess'' from the triple
power law fit at the 1.5~$\sigma$ level.

\begin{figure}
\epsscale{1.0}
\plotone{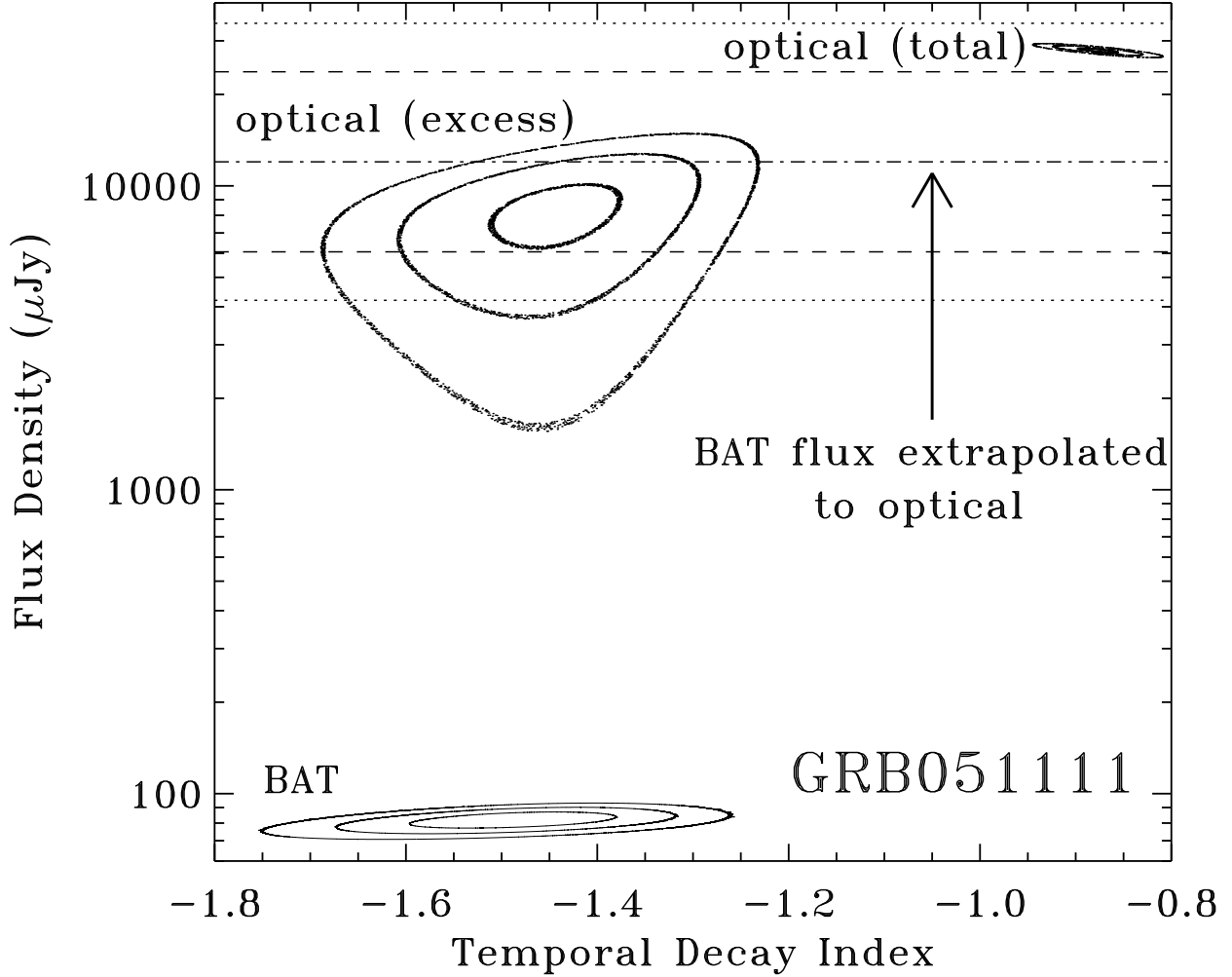}
\caption{ Comparing GRB\,051111's prompt optical and $\gamma$-ray flux
density and lightcurve evolution. Flux density is at the start of
ROTSE observations, 31.9~sec after the $\gamma$-ray onset. The
lightcurve evolution is measured as a power law index, $\alpha$, for
$f_{\nu}\propto t^{\alpha}$. The total optical lightcurve is fit from
31.9--150~sec and the $\gamma$-ray BAT lightcurve is fit from
15--200~sec. The optical excess is implied by the triple power law,
with a shallow prompt phase (Table \ref{tab:fits} and
Fig.~\ref{fig:1111_opt}). The excess' implied $\alpha$ matches the
$\gamma$-ray lightcurve. The excess is estimated by fitting the
optical ($t<$~1 ksec) and $\gamma$-ray (15--200~sec) simultaneously,
with an optical ``afterglow'' power law plus an excess constrained to
have the same $\alpha$ as the BAT data (see
\S\ref{1111prompt}). This optical excess fits the data well, showing a
good match to the $\gamma$-ray $\alpha$ determined from the BAT data
alone. The total optical lightcurve's index $\alpha$ is not a good
match. The excess' flux density level is consistent with an unbroken
spectral extrapolation of the BAT flux. The dot-dashed line shows the
best estimate of the BAT flux density at 31.9 sec, extrapolated to
optical frequencies via the photon index, $\Gamma$, (fit at $t>10$
sec, \S\ref{results:hifeatures}). The dashed lines give the
extrapolation range for 68\% confidence limits on $\Gamma$, and dotted
lines the 90\% range. The prompt $\gamma$-ray emission is compatible
with an unbroken extension to optical frequencies, producing the early
``excess'' optical component. See \S\,\ref{1111context} for a
comparison with other cases of prompt optical emission and their
optical-to-$\gamma$ spectra.}\label{fig:1111contour}
\end{figure}

We compare both the total optical flux and the estimated excess
optical flux to the $\gamma$-ray spectral extrapolation. The total
optical flux density at the early time is a good match for a simple
extension of the $\gamma$-ray power law spectrum. We determine
$\beta_{OPT - \gamma} = -0.554\pm0.005$ at 31.9~sec, while
$\beta_{\gamma} = -0.48\pm0.07$. These are compatible; the optical
flux at this time could be produced by the low-frequency tail of the
GRB. Two elements argue against this interpretation. First, it would
require a sudden change from optical flux entirely due to the GRB
component during the first exposure, to a similar optical flux
entirely due to the afterglow seconds later.  This is not a reasonable
model. Secondly, the total optical decay rate is significantly
shallower than the $\gamma$-ray decay as discussed above.
Thus we compare the flux densities of the optical excess and the
$\gamma$-rays and find the spectral index between them, $\beta =
-0.44\pm0.03$, is also compatible with $\beta_{\gamma}$. The optical
excess, considering both its temporal decay ($\alpha$) and its flux
level, could be produced by a spectrally unbroken low-frequency
extension of the GRB. This is demonstrated in
Fig.\,\ref{fig:1111contour}, which compares uncertainty contours for
flux densities and $\alpha$.

In conclusion, the prompt optical data of GRB\,051111 has an excess
over later optical evolution. The ``excess'' optical flux is
consistent with an extra component from the GRB emission. In this
case, the optical emission from the ``GRB proper'' is a simple
extension of the GRB spectrum from the BAT band of 15--150~keV, and
the spectral index from the optical excess to the $\gamma$-rays is
consistent with the spectral index within the $\gamma$-ray band. This
is unexpected, as there must eventually be a GRB spectral rollover at
low frequencies.

\subsection{Comparison of Prompt Detections}\label{1111context}

There have now been several cases with optical emission detected
contemporaneously with the $\gamma$-rays: GRB\,990123, GRB\,041219A,
GRB\,050319 (see Appendix\,\ref{1111tableexplain}), GRB\,050401, and
GRB\,050904, along with GRB\,051111 (discussed above) and GRB\,051109A
(as mentioned in \S\,\ref{results:opt2fl}, the first ROTSE point
overlaps with the tail end of GRB emission). There is no single
behavior among this group, spectrally or in lightcurve evolution.

The prompt optical emission of GRB\,990123 had an optical excess above
later afterglow evolution that was not correlated with the GRB peaks
\citep{abbbb99}, GRB\,041219A had optical emission correlated with the
GRB evolution \citep{vwwfs05,bbsfs05}, and GRB\,050401 had no
detectable excess prompt optical emission
\citep{rykaa05}. GRB\,050319 is similar to GRB\,050401 in that the
prompt optical detection (1st point) does not deviate from the
lightcurve \citep[see][]{qryaa05}. \citet{badgk06} discuss optical
emission during the very long, high-redshift GRB\,050904. They detect
optical flaring contemporaneously with X-ray flaring, at the end of
$\gamma$-ray emission, but do not discuss the optical comparison to
the $\gamma$-rays.

\begin{figure}
\epsscale{1.0}
\plotone{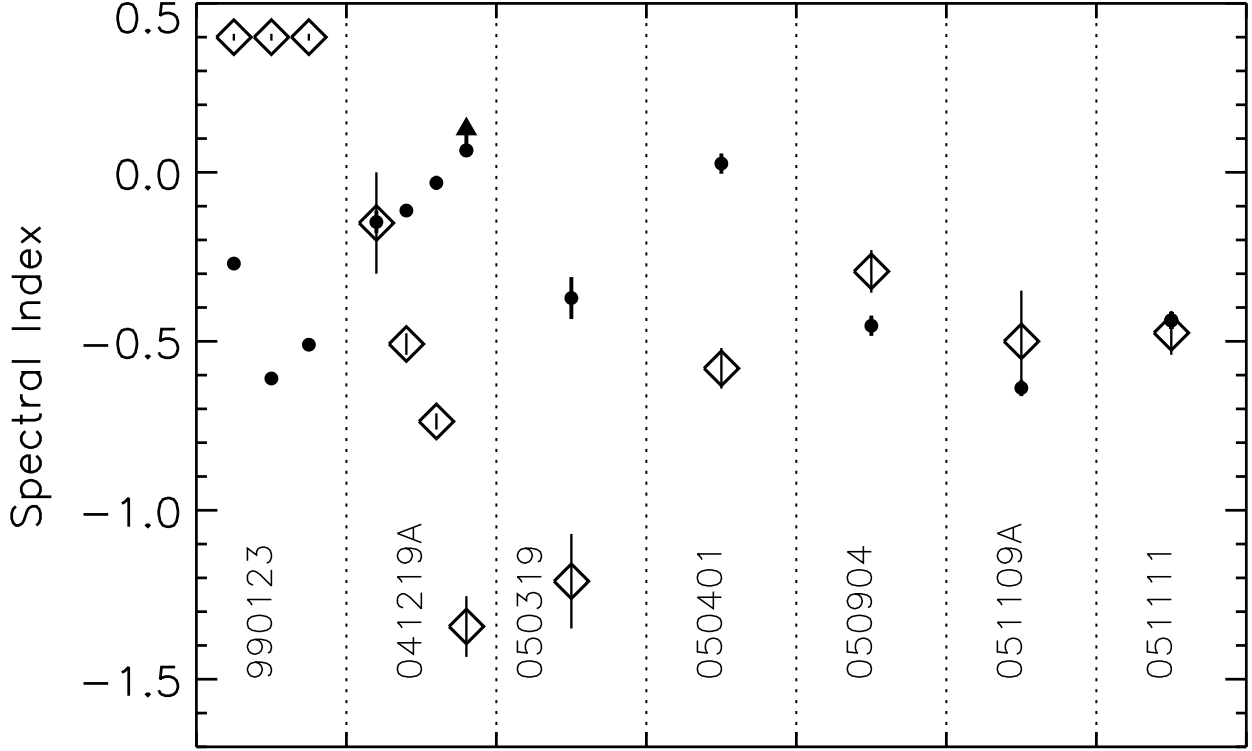}
\caption{ Comparison of prompt optical-to-$\gamma$-ray spectral
indices ($\beta_{OPT-\gamma}$) to the spectral index within the
$\gamma$-ray band ($\beta_{\gamma}$). The values in
Table\,\ref{tab:gamma} are plotted for each event, with small solid
circles for $\beta_{OPT-\gamma}$, and large open diamonds for
$\beta_{\gamma}$. Multiple measurements are for events with prompt
optical and $\gamma$-ray measurements during more than one time
interval. The sample shows all possible orderings of
$\beta_{OPT-\gamma}$ relative to $\beta_{\gamma}$. The GRB\,051111
$\beta_{OPT-\gamma}$ value uses the prompt optical ``excess'' component, not
the total optical flux. This component is consistent with an unbroken
spectral extrapolation from the high-energy GRB emission, see
\S\,\ref{1111prompt}. GRB\,051109A and GRB\,050904 are poorer candidates for
such an extension, as discussed in \S\,\ref{1111context}. Although
\citet{vwwfs05} show a correlation in the optical and $\gamma$-ray
lightcurves for GRB\,041219A, there must be a spectral break, such as a
synchrotron peak, between the two frequency bands.
}\label{fig:specindex}
\end{figure}

The optical-to-$\gamma$ spectral indices, $\beta_{OPT-\gamma}$, and
$\gamma$-ray band spectral indices, $\beta_{\gamma}$, are summarized
in Table \ref{tab:gamma} and plotted in
Figure\,\ref{fig:specindex}. Appendix\,\ref{1111tableexplain}
describes the sources of this information. The indices are compared to
see if any other events could have prompt optical emission as an
unbroken spectral extension of the $\gamma$-rays. In addition to
GRB\,051111, the GRB\,050904 event is a good candidate for such a
component. There are other possible examples, but the spectral
constraints are considerably poorer. GRB\,051109A is compatible with a
prompt optical extension, but the spectrum in the $\gamma$-rays has
significant uncertainty. The first time interval considered for the
broadband comparison in GRB\,041219A may be compatible; the
$\gamma$-ray spectral shape at the low end of the $\gamma$-ray band is
poorly constrained.

GRB\,050904 was a very long (in our reference frame) high-$z$ GRB. The
TAROT observations of \citet{badgk06} have an initial upper limit, two
constant detections, a flare, then upper limits. Only the first two
optical observations, up to 254 sec, have significant BAT flux
\citep[the 90\% flux duration is $225\pm10$\,sec,][]{sbbch05}.
The second observation (the first optical detection) is used to
get $\beta_{OPT-\gamma}$. This spectral index is compatible with an
extension of the BAT photon index (measured during the optical
observation from 169 to 254\,sec post-onset). However, the BAT flux
fades away by the next optical observation, and the
\citet{badgk06} optical flux does not. The two components may not be
from the same emission source.

Thus GRB\,051111 may be unusual, with a prompt optical component
compatible with the interpretation of a simple spectral extrapolation
from the $\gamma$-rays. There are several cases of prompt optical
observations, and no dominant behavior in the relative
optical/$\gamma$ comparisons. 

Beyond the extrapolation of $\beta_{\gamma}$ to the optical, the
comparison of $\beta_{OPT-\gamma}$ and $\beta_{\gamma}$ are not always
compatible with a single prompt synchrotron spectrum.  In the case of
GRB\,990123, the prompt optical flux is well above the $\gamma$-ray
spectral extrapolation and $\beta_{OPT-\gamma}$ is much softer than
$\beta_{\gamma}$. Connecting them requires a ``valley'' not seen in
the synchrotron spectrum. This may also be the case for GRB\,051109A,
although the uncertainties in the indices are too great to make this
determination.  In contrast, the GRB\,041219A, GRB\,050319, and
GRB\,050401 optical flux is well below the $\gamma$-ray spectral
extrapolation. These cases would be compatible with a synchrotron
spectrum having its flux density peak between the optical and
$\gamma$-ray frequencies. GRB\,041219A \citep[with an optical/$\gamma$
lightcurve correlation,][]{vwwfs05} would imply a prompt $p \approx
2$--2.7 for $\beta_{\gamma} = (1-p)/2$ (or $-p/2$ for the last time
interval tabled). For the cases of GRB\,050319 and GRB\,050401, there
is no lightcurve correlation and $\beta_{OPT-\gamma}$ with
$\beta_{\gamma}$ would imply a synchrotron peak at $\approx 0.5$,
$3$\,keV, respectively.

\subsection{Implications}

The GRB\,051109A and GRB\,051111 events continue to confirm what has
been previously noted in prompt comparisons: that there can often be a
prompt optical afterglow component distinct from the low-energy
emission tail of the GRB. In both cases, the prompt optical emission
smoothly connects to the later afterglow (albeit with an excess
component for GRB\,051111). In the context of the fireball model, this
would appear to indicate that the deceleration and thus the
establishment of the external shock occurs earlier than the end of
high-energy emission.  A separate afterglow component distinct from
the prompt emission is also implied by the likely interpretation of
\citet{nkgpg06}'s standard XRT afterglow shape. The initial fast
decline may be high-latitude emission from the end of the prompt GRB
emission component, superimposed on the shallower, distinct afterglow
component.

The existence of separate prompt and afterglow components is relevant
to the question of GRB emission models. The ``external--external''
shock model \citep{mr93} posits that the GRB's $\gamma$-ray emission is
produced by the fireball's external shock, and not by other means,
such as internal shocks within the relativistic flow. GRB variability
would be due to interactions with a clumpy external medium. This model
has been proposed as an explanation for GRBs with a small number of
$\gamma$-ray lightcurve peaks, such as GRB\,991216 \citep{rbcfx02},
and GRB\,970508 \citep[one of 10 GRBs with simple lightcurves analyzed
by][]{mkp04}. \citet{mkp04} examined the afterglow fits and
extrapolated the results to fluxes at the prompt GRB band. With a
single external shock, the afterglow would connect directly to the
GRB. Despite GRB\,051111's simple BAT lightcurve (see
Fig.\,\ref{fig:1111_gam}), it does not fit the external--external
picture. It has two components during the GRB -- a prompt optical
excess and the already-established afterglow. Its optical decay
changes after the end of GRB emission and does not extend
from the ``excess'' component apparently connected to the $\gamma$-ray
emission.

\section{Conclusions}\label{sec:conclusions}

GRB\,051109A and GRB\,051111 are two of the best-sampled cases to
analyze broadband comparisons of the prompt and very early post-burst
optical lightcurves to higher energy emission. The results continue to
show that there are diverse processes occurring during the early
afterglow phase.

GRB\,051109A has a break in both the optical and the X-ray near 0.5
day post-burst. It is consistent with being an achromatic transition,
although the X-ray data sampling does not permit this to be firmly
established. The breaks are shallower than expected for a jet break,
and are most easily explained by the cessation of steady energy
injection into the afterglow forward shock. The initial establishment
of the afterglow may be more complex at high energies (X-ray) than at
low ones (optical).

GRB\,051111's optical lightcurve decays more steeply during the prompt
emission than after the end of $\gamma$-ray detection. This indicates
a prompt excess above the continuing afterglow emission. Given the
temporal coincidence of the excess with the GRB emission, and its flux
level compatible with a spectral extrapolation of the $\gamma$-ray
flux, we interpret the excess as emission connected to the GRB. We do
not consider a reverse shock interpretation as likely. A separate
component for prompt emission, disconnected with the ongoing afterglow
emission, is not compatible with the external--external shock model
for this single-peaked $\gamma$-ray event.

In comparison to other prompt detections, the GRB\,051111 optical
component is unusual. The GRB\,051109A event may show optical flux at a
level compatible with a direct extrapolation of the $\gamma$-ray flux,
but it is not as well constrained. In the GRB\,041219A event, with an
optical lightcurve correlated with the GRB emission \citep{vwwfs05},
the flux level cannot be simply extrapolated. A spectral break is
implied between the optical and $\gamma$-ray frequencies.

In both GRB\,051109A and GRB\,051111, the afterglow emission is
ongoing during the prompt $\gamma$-ray emission. In these cases the
deceleration time to establish the external shock expected to power
the afterglow is shorter than the GRB duration. This is in contrast to
other cases where the afterglow is rising throughout or after the
prompt emission \citep[GRB\,050820A, GRB060605,][
respectively]{vwwag06,srsq06}. There are a variety of apparent
afterglow rise times, thus models of the GRB event and progenitor
environment must be capable to explaining such diverse results.

At tens of seconds after the GRB onset, the emitters are
ultrarelativistic and near the progenitor environment ($\sim$
light-week). GRB studies continue to uncover evidence of a wide
variety of processes underlying the emission during this early
phase. There is optical emission of the ``afterglow'' type during the
burst in both cases presented here.  With present capabilities it may
only be rarely possible to observe the ``afterglow onset'' (rise of
the forward shock) if it is usually well-established during the GRB
itself. The dearth of reverse shock signatures and evidence of steady
energy input for up to hours post-burst are clues to dynamic processes
at the heart of the collapse of massive stars. We are learning that
simple calculations are insufficient to address such data. It is
difficult to disentangle all source and environmental effects in order
to study the ultrarelativistic emission. The most promising avenue of
study uses prompt and early simultaneous observations at widely
separated frequencies. This coverage shows evidence that in
some events, such as GRB\,051109A, different frequencies resolve
themselves to the steady afterglow flux declines over different
timescales. Such observations may shed light on the early emission.

\acknowledgements

ROTSE-III has been supported by NASA grant NNG-04WC41G, NSF grant
AST-0407061, the Australian Research Council, the University of New
South Wales, the University of Texas, and the University of Michigan.
Work performed at LANL is supported through internal LDRD funding.
Special thanks to the observatory staff at McDonald Observatory,
especially David Doss.  

XRT work is supported at Observatorio Astronomica di Brera and
INAF-IASF Palermo by ASI grant I/R/039/04

The MDM work has been supported by the
National Science Foundation under grant 0206051 to JPH. 

The US Air Force provided the telescope time for RUCCD observations,
as well as on-site support and 80\% of the research funds for the
AFOSR and NSF jointly sponsored research, under grant number NSF
AST-0123487. The smooth installation and operation of the RUCCD has
been made possible through the extensive efforts and cooperation of
the staff at the telescope, and we would like to thank everyone who
has spent time helping us. Particular thanks for their help with our
project go to John Africano, Spence Ah You, Paul Bellaire, Peter
Figgis, John Flam, Tom Goggia, Nancy Ichikawa, Paul Kervin, Kevin Moore, Thomas
Nakagawa, Karl Rehder, Lewis Roberts, David Talent, and Darius Vunck.

The Digitized Sky Surveys were produced at the Space Telescope Science
Institute under U.S. Government grant NAG W-2166. The images of these
surveys are based on photographic data obtained using the Oschin
Schmidt Telescope on Palomar Mountain and the UK Schmidt
Telescope. The plates were processed into the present compressed
digital form with the permission of these institutions.

\appendix

\section{Data for Other Prompt Detections}\label{1111tableexplain}

In order to compare the GRB\,051111 prompt optical detection with other
cases, the spectral information for each event must be extracted in a
consistent fashion. Thus to derive Table\,\ref{tab:gamma}, for each
event we revisit the determination of the spectral index within the
$\gamma$-ray band, as well as between the optical and $\gamma$-rays.

The table gives the available simultaneous optical and $\gamma$-ray
detections for bursts to date. The comparisons are based on spectral
flux density, so both optical magnitudes and $\gamma$-ray fluxes are
converted to flux density at a particular frequency. The table
highlights whether the optical flux is above or below the spectrum
extrapolated from the $\gamma$-ray band by comparing
$\beta_{OPT-\gamma}$ to $\beta_{\gamma}$. As such, the $\gamma$-ray
frequency and spectral shape ($\beta_{\gamma}$) are for the {\it
lowest} well-measured $\gamma$-ray energy. In most cases, it is simply
the weighted average frequency across the energy band, given the
$\gamma$-ray spectral index. When possible, the optical to
$\gamma$-ray comparisons are made for more than one time interval during
the prompt overlap.

The following are details concerning each burst individually. 

For GRB\,990123, the data are from \citet{abbbb99}, Table 1 and
\citet{bbkpk99}, Table 2. The optical-to-$\gamma$ ratios of
\citet{bbkpk99} are adjusted to correspond to the final optical values
of \citet{abbbb99} (corrected for 0.05 mags of extinction), rather
than the GCN preliminary values. These subsequently are used to
produce $\beta_{OPT-\gamma}$. No uncertainties were provided for the
$\gamma$-ray flux densities, so Table\,\ref{tab:gamma} reports 3
significant figures, as in the source paper. The value of
$\beta_{\gamma}$ is taken from the Band model fits of \citet{bbkpk99}
for the entire event.

For GRB\,041219A, the data are from \citet{vwwfs05} (using its Fig. 4),
which corrects the optical flux for 4.9 mags of extinction. The $\gamma$-ray
frequency used, $\nu_{\gamma}$, is for the lowest energy of the four
$\gamma$-ray bins. $\beta_{\gamma}$ is fit for each time interval
using the four $\gamma$-ray frequencies and flux densities.  The first
time interval has a $\gamma$-ray spectrum that is not well-fit by a
single spectral index ($\chi^2=13$ for 2 degrees of freedom). It has
two entries in Table\,\ref{tab:gamma}. The first entry
(Table\,\ref{tab:gamma}, line 4) uses the overall least-squares fit
$\beta_{\gamma}$ for this time interval, despite the poor fit. The
second entry (Table\,\ref{tab:gamma}, line 5) uses $\beta_{\gamma}$
from the two lowest-energy $\gamma$-ray frequency bins, which has a
high uncertainty. The fourth time interval (Table\,\ref{tab:gamma},
line 8) has only upper limits for the flux densities in the optical
and the highest-energy $\gamma$ ray frequency bin. The spectral index
$\beta_{\gamma}$ is fit from only the first 3 frequency bins.

The GRB\,050319 optical point is from \citet{qryaa05}, corrected for
0.03 mags of extinction \citep{sfd98}. The $\gamma$-ray data are from
the {\it Swift} archive, analyzed with BATTOOLS to produce a flux
density and $\beta_{\gamma}$. This burst had more than one peak of
emission, and the soft $\gamma$-ray spectral index quoted is derived
from the final peak, from 130--170 sec post-onset. The optical
observations were taken after the end of the initially reported
$\gamma$-ray duration \citep{ksbbc05}. However, high-energy emission
was still detected in the count rate lightcurve. The faint emission
did not have high signal to noise during the 5-sec optical
observation. The $\gamma$-ray flux at the optical detection time was
estimated by interpolating a power law from the nearest 3 count rate
data points. The result was consistent with the power law
interpolation of the nearest 4 $\gamma$-ray detections in the tail of
the $\gamma$-ray peak, as well as with linear interpolations using
these 3 or 4 points. The signal-to-noise on the $\gamma$-ray count
rate detection is approximately 7; the low apparent signal-to-noise of
the flux density reported is due to uncertainty in the count rate
conversion to flux density.

The GRB\,050401 data is from \citet{rykaa05}. The photon index is
converted to $\beta_{\gamma}$. The optical flux density from
\citet{rykaa05}, Table 1 is corrected for 0.174 mags of extinction as
per \citet{sfd98} at the coordinates given in the paper. The
$\gamma$-ray flux density uses the complete 15--350 keV band,
converting to a flux density at 140 keV using $\beta_{\gamma}$.

The GRB\,050904 optical data is from \citet{badgk06}, corrected for
0.117 mags of extinction \citep[at the OT coordinates, as
  per][]{sfd98} and converted to mJy. $\gamma$-ray data was taken from
the {\it Swift} archive for this event. We used the standard BATTOOLS
presciption to determine $F_{\nu}$ and $\beta_{\gamma}$. We separated
the data into two response files (during and post-slew for this
interval), and combined them in XSPEC. As there is sufficient signal
for a good spectral extraction over the optical overlap
$t_{start}$--$t_{end}$, $\beta_{\gamma}$ is from this time interval
only. The asymmetric error bars are a result of this extraction. The
uncertainty in $\beta_{OPT-\gamma}$ compares the maximum optical and
minimum $\gamma$-ray (and vice-versa) to calculate the range of
$\beta$. The optical overlap analyzed here is the second interval (T2)
of the \citet{badgk06} observations. However, the first interval had
only an optical limit. The third interval had a similar optical flux
to T2, but the $\gamma$-ray flux is almost undetectable, as the GRB
$T_{90}$ duration is $225\pm10$~sec \citep{sbbch05}. Therefore other
optical comparisons were not well constrained. The evolution of this
event is further discussed in \S\,\ref{1111context}.

The GRB\,051109A optical data is taken from Table \ref{tab:photometry},
with an extinction correction of 0.511 mags (see
\S\,\ref{results:opt2fl}). The $\gamma$-ray data are from the {\it
Swift} archive, analyzed with BATTOOLS to produce a flux density and
$\beta_{\gamma}$. The spectral index $\beta_{\gamma}$ is for the
entire burst duration; there is no evidence of spectral evolution when
the $\gamma$-ray data is divided into two time bins. The symmetric
$F_{\nu}(\nu_{\gamma})$ error bars are statistical; the asymmetric
ones are the additional uncertainty in the conversion of count rate to
flux. Similarly, the index $\beta_{OPT-\gamma}$ has asymmetric error
bars from the count rate conversion uncertainty in the $\gamma$-ray
flux density.

For GRB\,051111, the optical data is from Table \ref{tab:photometry},
de-extincted by 0.433 mags (see \S\,\ref{results:opt2fl}). The
$\gamma$-ray data are from the {\it Swift} archive, as detailed in
\S\,\ref{sec:reductions}. In brief, standard BATTOOLS and XSPEC11
analysis were employed to extract spectral information for the event as
a whole, as well as the early ($-5$ to 10~sec) and late (10 to 50~sec)
parts of the burst (roughly divided to halve the signal). The burst
softened from early to late; the photon index of the late part of the
burst was used to produce the comparison $\beta_{\gamma}$. The optical
and $\gamma$-ray flux densities are compared in two different ways at
the midpoint of the first optical point (31.9~sec). The $\gamma$-ray
flux density is derived from a fit to the BAT lightcurve for $t >
15$~sec, which is well-fit by a power law $t^{-1.50\pm0.07}$. The
first optical flux comparison is to the initial optical detection. The
second is to the ``excess'' optical flux above the later, shallower
optical decay after the end of the GRB. While it is nominally at
31.9~sec, the optical excess is derived from the lightcurve evolution.
Line 13 uses the actual optical observation from
Table\,\ref{tab:photometry}, de-extincted. Line 14 uses the estimated
``optical excess'' from \S\,\ref{1111prompt} to make the
$\beta_{OPT-\gamma}$ comparison.

\newcommand{\noopsort}[1]{} \newcommand{\printfirst}[2]{#1}
  \newcommand{\singleletter}[1]{#1} \newcommand{\switchargs}[2]{#2#1}

\begin{deluxetable}{lcrrc}
\tablewidth{0pt}
\tablecaption{Optical Photometry for GRB\,051109A\label{tab:photometry}}
\tabletypesize{\scriptsize}
\tablehead{
  \colhead{Telescope} &
  \colhead{Filter} &
  \colhead{$t_{\mathrm{start}}$ (s)} &
  \colhead{$t_{\mathrm{end}}$ (s)} &
  \colhead{Magnitude}
}
\startdata
ROTSE & $C_R$ &    37.2 &    42.2 & $  14.991 \pm    0.061$ \\
ROTSE & $C_R$ &    44.3 &    49.3 & $  14.998 \pm    0.062$ \\
ROTSE & $C_R$ &    51.4 &    56.4 & $  15.150 \pm    0.071$ \\
ROTSE & $C_R$ &    58.5 &    63.5 & $  15.200 \pm    0.070$ \\
ROTSE & $C_R$ &    65.6 &    70.6 & $  15.347 \pm    0.080$ \\
ROTSE & $C_R$ &    72.7 &    77.7 & $  15.306 \pm    0.079$ \\
ROTSE & $C_R$ &    79.8 &    84.8 & $  15.443 \pm    0.089$ \\
ROTSE & $C_R$ &    86.9 &    91.9 & $  15.478 \pm    0.091$ \\
ROTSE & $C_R$ &    94.1 &    99.1 & $  15.368 \pm    0.077$ \\
ROTSE & $C_R$ &   101.2 &   106.2 & $  15.530 \pm    0.092$ \\
ROTSE & $C_R$ &   119.5 &   139.5 & $  15.703 \pm    0.053$ \\
ROTSE & $C_R$ &   156.6 &   176.6 & $  15.899 \pm    0.062$ \\
ROTSE & $C_R$ &   186.2 &   206.2 & $  15.960 \pm    0.067$ \\
ROTSE & $C_R$ &   215.1 &   235.1 & $  15.916 \pm    0.059$ \\
ROTSE & $C_R$ &   245.4 &   265.4 & $  16.081 \pm    0.069$ \\
ROTSE & $C_R$ &   274.7 &   294.7 & $  16.208 \pm    0.079$ \\
ROTSE & $C_R$ &   303.8 &   323.8 & $  16.199 \pm    0.073$ \\
ROTSE & $C_R$ &   332.8 &   352.8 & $  16.476 \pm    0.096$ \\
ROTSE & $C_R$ &   361.9 &   381.9 & $   16.55 \pm     0.11$ \\
ROTSE & $C_R$ &   391.5 &   411.5 & $  16.502 \pm    0.098$ \\
ROTSE & $C_R$ &   421.0 &   481.0 & $  16.468 \pm    0.057$ \\
ROTSE & $C_R$ &   490.2 &   550.2 & $  16.753 \pm    0.070$ \\
ROTSE & $C_R$ &   559.3 &   619.3 & $  16.805 \pm    0.073$ \\
ROTSE & $C_R$ &   628.5 &   688.5 & $  16.987 \pm    0.094$ \\
ROTSE & $C_R$ &   697.6 &   757.6 & $  16.912 \pm    0.083$ \\
ROTSE & $C_R$ &   766.6 &   826.6 & $   17.06 \pm     0.10$ \\
ROTSE & $C_R$ &   835.7 &   895.7 & $  17.064 \pm    0.097$ \\
ROTSE & $C_R$ &   904.9 &   964.9 & $   17.37 \pm     0.14$ \\
ROTSE & $C_R$ &   974.0 &  1034 & $   17.59 \pm     0.16$ \\
ROTSE & $C_R$ &  1043 &  1103 & $   17.19 \pm     0.11$ \\
ROTSE & $C_R$ &  1113 &  1450 & $  17.411 \pm    0.069$ \\
ROTSE & $C_R$ &  1459 &  1796 & $  17.559 \pm    0.080$ \\
ROTSE & $C_R$ &  1805 &  2142 & $  17.863 \pm    0.095$ \\
ROTSE & $C_R$ &  2151 &  3179 & $  17.889 \pm    0.078$ \\
ROTSE & $C_R$ &  3188 &  4368 & $   18.29 \pm     0.14$ \\
ROTSE & $C_R$ &  4378 &  5406 & $   18.19 \pm     0.11$ \\
ROTSE & $C_R$ &  5415 &  7480 & $   18.31 \pm     0.11$ \\
ROTSE & $C_R$ &  7490 &  9558 & $   18.46 \pm     0.12$ \\
ROTSE & $C_R$ &  9567 & 12048 & $   18.57 \pm     0.13$ \\
ROTSE & $C_R$ & 12060 & 14540 & $   18.90 \pm     0.17$ \\
MDM & $r_R$ &    2135 &    2255 & $  18.186 \pm    0.048$ \\
MDM & $r_R$ &    2292 &    2412 & $  18.214 \pm    0.075$ \\
MDM & $r_R$ &    2448 &    2568 & $  18.333 \pm    0.060$ \\
MDM & $r_R$ &    2996 &    3296 & $  18.365 \pm    0.061$ \\
MDM & $r_R$ &    3332 &    3632 & $  18.546 \pm    0.067$ \\
MDM & $r_R$ &    3669 &    3969 & $  18.535 \pm    0.066$ \\
MDM & $r_R$ &    4005 &    4305 & $  18.486 \pm    0.081$ \\
MDM & $r_R$ &    4342 &    4642 & $  18.574 \pm    0.037$ \\
MDM & $r_R$ &    4678 &    4978 & $  18.747 \pm    0.051$ \\
MDM & $r_R$ &    5014 &    5314 & $  18.835 \pm    0.057$ \\
MDM & $r_R$ &    5351 &    5651 & $  18.858 \pm    0.071$ \\
MDM & $r_R$ &    5687 &    5987 & $  18.865 \pm    0.040$ \\
MDM & $r_R$ &    6023 &    6323 & $  18.931 \pm    0.041$ \\
MDM & $r_R$ &    6360 &    6660 & $  18.933 \pm    0.067$ \\
MDM & $r_R$ &    6696 &    6996 & $  18.874 \pm    0.074$ \\
MDM & $r_R$ &    7032 &    7332 & $  18.995 \pm    0.094$ \\
MDM & $r_R$ &    7368 &    7668 & $  19.160 \pm    0.086$ \\
MDM & $r_R$ &    7704 &    8004 & $  19.022 \pm    0.086$ \\
MDM & $r_R$ &    8377 &    8677 & $  19.093 \pm    0.066$ \\
MDM & $r_R$ &    8713 &    9013 & $  19.192 \pm    0.080$ \\
MDM & $r_R$ &    9049 &    9349 & $   19.31 \pm     0.11$ \\
MDM & $r_R$ &   10240 &   10840 & $  19.159 \pm    0.093$ \\
MDM & $r_R$ &   10880 &   12750 & $   19.27 \pm     0.17$ \\
MDM & $r_R$ &   89850 &   90450 & $  21.190 \pm    0.083$ \\
MDM & $r_R$ &  100600 &  101200 & $   21.44 \pm     0.10$ \\
MDM & $r_R$ &  264300 &  266100 & $   22.48 \pm     0.11$ \\
MDM & $R$ & 1.046$\times10^{6}$ & 1.049$\times10^{6}$ & $23.79 \pm  0.17$ \\
MDM & $R$ & 2.017$\times10^{7}$ & 2.017$\times10^{7}$ & $23.70 \pm  0.16$ \\
\enddata
\tablecomments{All times are in seconds since the burst onset, 01:12:15.5 UT (see \S\,\ref{sec:observations}). ROTSE $C_R$ magnitudes are for unfiltered CCD magnitudes referenced to $R$ with the USNOB 1.0 standards. MDM $r_R$ magnitudes are $r$-band observations that are referenced to $R$ using two USNOB 1.0 $R$ standards.}
\end{deluxetable}

\begin{deluxetable}{lcrrc}
\tablewidth{0pt}
\tablecaption{Optical Photometry for GRB\,051111\label{tab:photometry2}}
\tabletypesize{\scriptsize}
\tablehead{
  \colhead{Instrument} &
  \colhead{Filter} &
  \colhead{$t_{\mathrm{start}}$ (s)} &
  \colhead{$t_{\mathrm{end}}$ (s)} &
  \colhead{Magnitude}
}
\startdata
ROTSE & $C_R$ &    29.4 &    34.4 & $  13.062 \pm    0.029$ \\
ROTSE & $C_R$ &    36.5 &    41.5 & $  13.262 \pm    0.029$ \\
ROTSE & $C_R$ &    43.6 &    48.6 & $  13.372 \pm    0.028$ \\
ROTSE & $C_R$ &    50.7 &    55.7 & $  13.512 \pm    0.032$ \\
ROTSE & $C_R$ &    57.8 &    62.8 & $  13.610 \pm    0.033$ \\
ROTSE & $C_R$ &    64.9 &    69.9 & $  13.753 \pm    0.037$ \\
ROTSE & $C_R$ &    72.0 &    77.0 & $  13.798 \pm    0.038$ \\
ROTSE & $C_R$ &    79.1 &    84.1 & $  13.908 \pm    0.039$ \\
ROTSE & $C_R$ &    86.3 &    91.3 & $  14.049 \pm    0.042$ \\
ROTSE & $C_R$ &    93.4 &    98.4 & $  14.068 \pm    0.036$ \\
ROTSE & $C_R$ &   111.4 &   131.4 & $  14.352 \pm    0.028$ \\
ROTSE & $C_R$ &   148.9 &   168.9 & $  14.547 \pm    0.031$ \\
ROTSE & $C_R$ &   177.8 &   197.8 & $  14.683 \pm    0.027$ \\
ROTSE & $C_R$ &   207.0 &   227.0 & $  14.751 \pm    0.034$ \\
ROTSE & $C_R$ &   236.8 &   256.8 & $  14.885 \pm    0.057$ \\
ROTSE & $C_R$ &   265.8 &   285.8 & $  14.986 \pm    0.049$ \\
ROTSE & $C_R$ &   295.3 &   315.3 & $  14.993 \pm    0.047$ \\
ROTSE & $C_R$ &   324.9 &   344.9 & $  15.097 \pm    0.045$ \\
ROTSE & $C_R$ &   353.9 &   373.9 & $  15.156 \pm    0.075$ \\
ROTSE & $C_R$ &   383.5 &   403.5 & $  15.211 \pm    0.049$ \\
ROTSE & $C_R$ &   412.9 &   432.9 & $  15.181 \pm    0.063$ \\
ROTSE & $C_R$ &   441.9 &   461.9 & $  15.271 \pm    0.075$ \\
ROTSE & $C_R$ &   471.0 &   491.0 & $  15.346 \pm    0.084$ \\
ROTSE & $C_R$ &   500.3 &   520.3 & $  15.591 \pm    0.099$ \\
ROTSE & $C_R$ &   529.3 &   549.3 & $  15.601 \pm    0.075$ \\
ROTSE & $C_R$ &   558.2 &   578.2 & $  15.641 \pm    0.085$ \\
ROTSE & $C_R$ &   587.8 &   607.8 & $  15.487 \pm    0.060$ \\
ROTSE & $C_R$ &   616.8 &   636.8 & $  15.577 \pm    0.050$ \\
ROTSE & $C_R$ &   645.8 &   665.8 & $  15.707 \pm    0.072$ \\
ROTSE & $C_R$ &   674.9 &   694.9 & $  15.825 \pm    0.070$ \\
ROTSE & $C_R$ &   703.9 &   723.9 & $  15.844 \pm    0.079$ \\
ROTSE & $C_R$ &   733.1 &   753.1 & $   15.78 \pm     0.11$ \\
ROTSE & $C_R$ &   762.6 &   782.6 & $   15.82 \pm     0.11$ \\
ROTSE & $C_R$ &   791.7 &   811.7 & $  15.934 \pm    0.088$ \\
ROTSE & $C_R$ &   821.2 &   841.2 & $  15.897 \pm    0.085$ \\
ROTSE & $C_R$ &   850.4 &   870.4 & $  15.829 \pm    0.078$ \\
ROTSE & $C_R$ &   879.9 &   899.9 & $  15.922 \pm    0.068$ \\
ROTSE & $C_R$ &   909.5 &   929.5 & $   15.90 \pm     0.15$ \\
ROTSE & $C_R$ &   938.4 &   958.4 & $   16.03 \pm     0.13$ \\
ROTSE & $C_R$ &   967.8 &   987.8 & $  16.059 \pm    0.089$ \\
ROTSE & $C_R$ &   996.9 &  1016.9 & $   16.13 \pm     0.13$ \\
ROTSE & $C_R$ &  1026 &  1046 & $   16.01 \pm     0.12$ \\
ROTSE & $C_R$ &  1055 &  1075 & $   16.12 \pm     0.10$ \\
ROTSE & $C_R$ &  1084 &  1104 & $   16.08 \pm     0.11$ \\
ROTSE & $C_R$ &  1114 &  1134 & $  16.143 \pm    0.099$ \\
ROTSE & $C_R$ &  1143 &  1163 & $   16.22 \pm     0.11$ \\
ROTSE & $C_R$ &  1173 &  1193 & $   16.30 \pm     0.11$ \\
ROTSE & $C_R$ &  1202 &  1222 & $   16.15 \pm     0.13$ \\
ROTSE & $C_R$ &  1231 &  1251 & $   16.31 \pm     0.12$ \\
ROTSE & $C_R$ &  1261 &  1281 & $   16.25 \pm     0.11$ \\
ROTSE & $C_R$ &  1290 &  1310 & $   16.29 \pm     0.13$ \\
ROTSE & $C_R$ &  1319 &  1339 & $   16.34 \pm     0.14$ \\
ROTSE & $C_R$ &  1348 &  1368 & $   16.33 \pm     0.12$ \\
ROTSE & $C_R$ &  1378 &  1398 & $   16.39 \pm     0.14$ \\
ROTSE & $C_R$ &  1407 &  1427 & $   16.42 \pm     0.14$ \\
ROTSE & $C_R$ &  1436 &  1456 & $   16.46 \pm     0.14$ \\
ROTSE & $C_R$ &  1466 &  1486 & $   16.78 \pm     0.21$ \\
ROTSE & $C_R$ &  1495 &  1515 & $   16.46 \pm     0.17$ \\
ROTSE & $C_R$ &  1524 &  1544 & $  16.226 \pm    0.089$ \\
ROTSE & $C_R$ &  1554 &  1574 & $   16.56 \pm     0.16$ \\
ROTSE & $C_R$ &  1583 &  1603 & $   16.86 \pm     0.19$ \\
ROTSE & $C_R$ &  1613 &  1633 & $   16.66 \pm     0.24$ \\
ROTSE & $C_R$ &  1642 &  1662 & $   16.61 \pm     0.15$ \\
ROTSE & $C_R$ &  1671 &  1691 & $   16.66 \pm     0.26$ \\
ROTSE & $C_R$ &  1700 &  1720 & $   16.80 \pm     0.20$ \\
ROTSE & $C_R$ &  1729 &  1749 & $   16.68 \pm     0.19$ \\
ROTSE & $C_R$ &  1758 &  1778 & $   16.68 \pm     0.16$ \\
ROTSE & $C_R$ &  1788 &  1808 & $   16.76 \pm     0.18$ \\
ROTSE & $C_R$ &  1817 &  1837 & $ > 17.1$ \\
ROTSE & $C_R$ &  1846 &  1866 & $   16.90 \pm     0.16$ \\
ROTSE & $C_R$ &  1953 &  1973 & $   16.68 \pm     0.20$ \\
ROTSE & $C_R$ &  1983 &  2003 & $   16.72 \pm     0.27$ \\
ROTSE & $C_R$ &  2012 &  2032 & $   16.87 \pm     0.23$ \\
ROTSE & $C_R$ &  2042 &  2062 & $   16.60 \pm     0.11$ \\
ROTSE & $C_R$ &  2071 &  2091 & $   16.64 \pm     0.17$ \\
ROTSE & $C_R$ &  2100 &  2237 & $  16.816 \pm    0.084$ \\
ROTSE & $C_R$ &  2246 &  2383 & $   17.04 \pm     0.16$ \\
ROTSE & $C_R$ &  2392 &  2529 & $   17.05 \pm     0.11$ \\
ROTSE & $C_R$ &  2538 &  2675 & $   17.22 \pm     0.10$ \\
ROTSE & $C_R$ &  2684 &  2821 & $   17.34 \pm     0.15$ \\
ROTSE & $C_R$ &  2830 &  2968 & $  17.315 \pm    0.096$ \\
ROTSE & $C_R$ &  2977 &  3407 & $   17.30 \pm     0.11$ \\
ROTSE & $C_R$ &  3417 &  3845 & $   17.69 \pm     0.21$ \\
ROTSE & $C_R$ &  3855 &  4283 & $   17.74 \pm     0.16$ \\
ROTSE & $C_R$ &  4293 &  4721 & $   17.86 \pm     0.19$ \\
ROTSE & $C_R$ &  4730 &  5730 & $   18.24 \pm     0.12$ \\
ROTSE & $C_R$ &  5739 &  6604 & $   18.43 \pm     0.24$ \\
ROTSE & $C_R$ &  6613 &  8561 & $   18.96 \pm     0.37$ \\
RUCCD & $V$ & 2972\tablenotemark{a} & 4104 & $17.1\pm0.3$ \\
RUCCD & $R$ & 1908\tablenotemark{b} & 2520 & $16.7\pm0.4$  \\
RUCCD & $R$ & 2972\tablenotemark{c} & 4104 & $17.5\pm0.2$  \\
RUCCD & $I$ & 1908\tablenotemark{d} & 2520 & $15.7\pm0.3$  \\
RUCCD & $I$ & 2972\tablenotemark{e} & 4104 & $16.9\pm0.2$ \\
MDM & $R$ & 87900 & 91400 & $21.63\pm0.10$ \\
\enddata
\tablenotetext{a}{RUCCD observations were taken by shifting through the $VRI$ filters during this interval. The mean time of the coadded images used for this point is 3488~sec.}
\tablenotetext{b}{RUCCD observations were taken by shifting through the $RI$ filters during this interval. The mean time of the coadded images used for this point is 2350~sec.}
\tablenotetext{c}{RUCCD observations were taken by shifting through the $VRI$ filters during this interval. The mean time of the coadded images used for this point is 3794~sec.}
\tablenotetext{d}{RUCCD observations were taken by shifting through the $RI$ filters during this interval. The mean time of the coadded images used for this point is 2199~sec.}
\tablenotetext{e}{RUCCD observations were taken by shifting through the $VRI$ filters during this interval. The mean time of the coadded images used for this point is 4006~sec.}
\tablecomments{All times are in seconds since the burst onset, 05:59:39 UT (see \S\,\ref{sec:observations}). ROTSE $C_R$ magnitudes are for unfiltered CCD magnitudes referenced to $R$ with the USNOB 1.0 standards.}
\end{deluxetable}

\begin{deluxetable}{llc}
\tablewidth{0pt}
\tablecaption{XRT Fluxes for GRB\,051109A\label{tab:XRT}}
\tabletypesize{\scriptsize}
\tablehead{
  \colhead{$t_{\mathrm{start}}$ (ksec)} &
  \colhead{$t_{\mathrm{end}}$ (ksec)} &
  \colhead{0.2--10 keV Flux} \\
  \colhead{} &
  \colhead{} &
  \colhead{($10^{-12}$erg\,cm$^{-2}$s$^{-1}$)} \\
}
\startdata
0.1325 & 0.1355 &  1380$\pm$190 \\
0.1355 & 0.1385 &  1550$\pm$210 \\
0.1385 & 0.1415 &  1280$\pm$190 \\
0.1415 & 0.1445 & 810$\pm$150 \\
0.1445 & 0.1475 & 950$\pm$170 \\
0.1475 & 0.1505 & 920$\pm$160 \\
0.1505 & 0.1535 & 950$\pm$160 \\
0.1535 & 0.1565 & 570$\pm$130 \\
0.1565 & 0.1595 & 710$\pm$140 \\
0.1595 & 0.1625 & 810$\pm$150 \\
0.1625 & 0.1685 & 770$\pm$100 \\
0.1685 & 0.1745 & 638$\pm$95 \\
0.1745 & 0.1805 & 543$\pm$90 \\
0.1805 & 0.1865 & 434$\pm$81 \\
0.1865 & 0.1925 & 475$\pm$85 \\
0.1925 & 0.1985 & 461$\pm$79 \\
0.1985 & 0.2045 & 240$\pm$120 \\
 3.432 &  3.492 &  100$\pm$19 \\
 3.492 &  3.552 & 59$\pm$18 \\
 3.552 &  3.612 & 69$\pm$17 \\
 3.565 &  3.685 & 93$\pm$16 \\
 3.612 &  3.672 &  142$\pm$33 \\
 3.685 &  3.805 & 84$\pm$11 \\
 3.805 &  3.925 & 74$\pm$11 \\
 3.925 &  4.045 & 78$\pm$11 \\
 4.045 &  4.165 & 63.7$\pm$9.8 \\
 4.165 &  4.285 & 66.8$\pm$9.7 \\
 4.285 &  4.405 & 68$\pm$10 \\
 4.405 &  4.525 & 86$\pm$11 \\
 4.525 &  4.645 & 59.4$\pm$9.7 \\
 4.645 &  4.765 & 68.3$\pm$9.7 \\
 4.765 &  4.885 & 52.0$\pm$8.6 \\
 4.885 &  5.005 & 59.7$\pm$9.5 \\
 5.005 &  5.125 & 53.0$\pm$9.0 \\
 5.125 &  5.245 & 47.6$\pm$8.6 \\
 5.245 &  5.365 & 60.9$\pm$9.7 \\
 5.365 &  5.485 & 51.4$\pm$9.1 \\
 5.485 &  5.605 & 60.9$\pm$9.7 \\
 5.605 &  5.845 & 51.2$\pm$6.1 \\
 5.845 &  6.085 & 53.8$\pm$8.7 \\
 9.205 &  9.445 & 46.0$\pm$6.1 \\
 9.445 &  9.685 & 24.3$\pm$4.4 \\
 9.685 &  9.925 & 28.9$\pm$4.9 \\
 9.925 & 10.165 & 29.6$\pm$4.9 \\
 10.17 &  10.41 &  25.9$\pm$4.4 \\
 10.41 &  10.65 &  26.5$\pm$4.5 \\
 10.65 &  10.89 &  24.8$\pm$4.7 \\
 10.89 &  11.13 &  29.4$\pm$4.6 \\
 11.13 &  11.37 &  33.9$\pm$5.0 \\
 11.37 &  11.61 &  24.8$\pm$4.7 \\
 11.61 &  11.85 &  26.1$\pm$5.7 \\
 14.92 &  15.31 &  16.6$\pm$2.8 \\
 15.31 &  15.70 &  17.7$\pm$2.1 \\
 15.70 &  16.09 &  16.9$\pm$2.0 \\
 16.09 &  16.48 &  15.4$\pm$2.0 \\
 16.48 &  16.87 &  16.6$\pm$2.1 \\
 16.87 &  17.26 &  14.0$\pm$1.9 \\
 17.26 &  17.65 &  11.4$\pm$2.0 \\
 20.77 &  21.16 &  10.3$\pm$2.0 \\
 21.16 &  21.55 &  14.7$\pm$1.9 \\
 21.55 &  21.94 &  13.5$\pm$1.9 \\
 21.94 &  22.33 &  11.3$\pm$1.8 \\
 22.33 &  22.72 &  12.5$\pm$1.8 \\
 22.72 &  23.11 &  14.7$\pm$1.9 \\
 23.11 &  23.50 &  14.5$\pm$2.5 \\
 26.62 &  27.01 &  10.6$\pm$2.7 \\
 50.59 &  63.08 & 4.14$\pm$0.30 \\
 78.67 &  84.90 & 3.40$\pm$0.38 \\
 84.94 & 167.08 & 1.290$\pm$0.099 \\
 172.6 &  254.6 &  0.65$\pm$0.12 \\
 258.8 &  283.5 &  0.472$\pm$0.076 \\
 350.7 &  376.1 &  0.382$\pm$0.076 \\
 437.5 &  461.7 &  0.239$\pm$0.059 \\
 605.1 &  665.3 &  0.163$\pm$0.030 \\
 687.4 &  855.3 &  0.161$\pm$0.027 \\
 860.0 & 1202 &  0.084$\pm$0.012 \\
1207 & 1550 &  0.074$\pm$0.011 \\
\enddata
\tablecomments{All times are in seconds since the burst onset, 01:12:15.5 UT (see \S\,\ref{sec:observations}).}
\end{deluxetable}

\begin{deluxetable}{lcccccccccc}
\rotate
\tablewidth{0pt}
\tablecaption{Power Law Fits\label{tab:fits}}
\tabletypesize{\scriptsize}
\tablehead{
  \colhead{Dataset} &
  \colhead{$t_{0}(s)$} &
  \colhead{$F_{0}$($\mu$Jy)\tablenotemark{a} } &
  \colhead{$\alpha_1$ } &
  \colhead{$t_{\mathrm{break}}1$(ksec)} &
  \colhead{$S$\tablenotemark{b}} &
  \colhead{$\alpha_2$ } &
  \colhead{$t_{\mathrm{break}}2$(ksec)} &
  \colhead{$S$\tablenotemark{b}} &
  \colhead{$\alpha_3$ } &
  \colhead{color term\tablenotemark{c}} 
}
\startdata
051109A optical\tablenotemark{d} & na & $50.6\pm6.5$ & $-0.6520\pm0.0082$ & $52.4\pm9.2$ & 50 & $-1.47\pm0.18$ & na & na & na & $1.513\pm0.043$\\
051109A XRT 1\tablenotemark{e} & 150 & $40.0\pm1.8$ & $-3.20\pm0.36$ & na & na & na & na & na & na & na \\
051109A XRT 2\tablenotemark{e} & na & $0.35\pm0.12$  &$-1.036\pm0.034$ & $34\pm10$ & 9 & $-1.32\pm0.032$ & na & na & na & na \\
051109A XRT\tablenotemark{e} & na & $18.7\pm3.9$  & $-3.28\pm0.49$ & $0.189\pm0.016$ & -9 & $-0.599\pm0.053$ & $6.59\pm0.62$ & 9 & $-1.237\pm0.017$ & na \\
051111 optical & na & $8500\pm1000$ & $-0.876\pm0.021$ & $0.124\pm0.018$ & $-50$ & $-0.742\pm0.013$ & $1.100\pm0.088$ & 50 & $-1.169\pm0.022$ & na \\
\enddata
\tablenotetext{a}{The normalization of a single power law is
$F_{0}(t/t_{0})^{\alpha}$, and $t_{0}$ is selected for convenience
within the data's time range. The double and triple power law fits
have a different normalization. The formula for a double power law is
$F_{0}(t/t_{\mathrm{break}}1)^{\alpha_1} (1 +
(t/t_{\mathrm{break}}1)^{S (\alpha_1 - \alpha_2)})^{-1/S}$. The
formula for a triple power law is $F_{0}(t/t_{\mathrm
break~1})^{\alpha_1} (1 + (t/t_{\mathrm{break}}1)^{S_1 (\alpha_1 -
\alpha_2)})^{-1/S_1} (1 + (t/t_{\mathrm{break}}2)^{S_2 (\alpha_2 -
\alpha_3)})^{-1/S_2}$.}
\tablenotetext{b}{The $S$ values are sharpness parameters for the
breaks (see note above). In no case was $S$ well-determined in the
fit, and fixed values are selected to produce sharp breaks.}
\tablenotetext{c}{Multiplicative factor applied to MDM $R$-equivalent data in a fit, relative to the unfiltered $R$-equivalent ROTSE flux densities.}
\tablenotetext{d}{The GRB\,051109A optical fit is to all points including the
last (host) one. A constant is added to the double power law fit model. The fitted host level is $1.40\pm0.21$ $\mu$Jy, within 2.5$\sigma$ of the measured host flux (corrected for extinction and the MDM color term). The MDM color term fit is dominated by the data during the MDM/ROTSE overlap, the $r_{R}$ observations. A fit in which only $r_{R}$ data get a color term (the last 2 MDM points left with no color term) does not affect the results.}
\tablenotetext{e}{The XRT observations are divided into the first
orbit (XRT 1) and all subsequent data (XRT 2). As seen in
Fig. \ref{fig:fig1109a}, there is a data gap and the early and late
evolution do not match. Thus there are three fits, the first orbit
(XRT 1), the later orbits (XRT 2), and an overall fit of all the data
(XRT)}
\tablecomments{Data for GRB\,051109A and GRB\,051111 optical taken
from Tables \ref{tab:photometry} and \ref{tab:photometry2}, corrected
for Galactic extinction, and converted to flux densities as discussed
in \S\,\ref{results:opt2fl}. GRB\,051109A XRT data from Table
\ref{tab:XRT}. Not every fit uses all the
parameters tabled; when a parameter was not applicable, this is
indicated in the table as ``na''. The values of $\chi^2$ and degrees
of freedom (DOF) for each fit are as follows: GRB\,051109A optical - 96
for 62 DOF, GRB\,051109A XRT 1 - 14 for 15 DOF, GRB\,051109A XRT 2 - 53
for 55 DOF, GRB\,051109A XRT - 70 for 71 DOF, and GRB\,051111 optical -
78 for 83 DOF.}
\end{deluxetable}

\begin{deluxetable}{lcccccccc}
\tablewidth{0pt}
\tablecaption{Gamma/Optical Comparisions\label{tab:gamma}}
\tabletypesize{\scriptsize}
\tablehead{
  \colhead{GRB} &
  \colhead{$t_{start}(s)$} &
  \colhead{$t_{end}(s)$} &
  \colhead{Band} &
  \colhead{$F_{\nu}(\nu_{OPT})$(mJy) } &
  \colhead{$\nu_{\gamma}$($10^{18}$Hz)} &
  \colhead{$F_{\nu}(\nu_{\gamma})$ ($\mu$Jy) } &
  \colhead{$\beta_{\gamma} = 1-\Gamma$} &
 % \colhead{test} 
\colhead{$\beta_{OPT-\gamma}$}
}
\startdata
990123 & 22.2 & 27.2 & $C_V$ & $79.6\pm5.1$ & $24$ & 4450  & $0.40\pm0.01$  & -0.270 \\
.. & 47.4 & 52.4 & .. & $1090\pm20$ & $24$ & 1630  & ..  & -0.609 \\
.. & 72.7 & 77.7 & .. & $392\pm11$ & $24$ & 1710  & ..  & -0.508 \\
041219A & 202.9 & 275.5 &  $C_R$ & $2.88 \pm 0.87$ & $5.0 \pm 1.3$ &  $738\pm39$ & $-0.39$ & $-0.147\pm0.033$ \\
041219A & .. & .. &  .. & .. & .. & .. & $-0.15 \pm 0.15$ & $-0.147\pm0.033$ \\
.. & 288.0 & 318.0 &  .. & $10.3\pm1.0$ & .. &$3600\pm190$  & $-0.508\pm0.032$ & $-0.113\pm0.012$\\
.. & 330.4 & 402.9 &  .. & $3.84\pm0.76$ & .. & $2882\pm75$ & $-0.737\pm0.024$ & $-0.031 \pm 0.021$\\
.. & 415.4 & 573.1 &  .. & $< 1.0$ & .. & $583\pm31$ &$-1.344 \pm 0.090$ & $> -0.065$\\
050319 & 162.8 & 167.8 & .. & $1.30\pm0.17$ & 13 & $29^{+18}_{-12}$ & $-1.21\pm0.14$ & $-0.372\pm0.062$ \\
050401 & 33.2 & 38.2 & .. & $0.69\pm0.19$ & 34 & $877\pm28$ & $-0.58\pm0.06$ & $0.026 \pm 0.030$ \\
050904 & 169.0 & 253.8 & $C_I$ & $19.4^{+5.4}_{-4.0}$ & 18 & $134.1^{+4.8}_{-12.1}$ & $-0.293 \pm 0.063$ & $-0.454^{+0.024}_{-0.031}$ \\
051109A & 37.7 & 42.7 & $C_R$ & $4.97\pm0.28$ & 17 & $6.10^{+0.12}_{-1.71}\pm1.47$ & $-0.50\pm0.15$ & $-0.638^{+0.002}_{-0.031}\pm0.024$ \\
051111 & 29.4 & 34.4 & .. & $27.34\pm0.74$ & 17 & $81.7\pm4.1$ & $-0.475\pm0.065$ & $-0.5539\pm0.0054$ \\
.. & 31.9 & 31.9 & .. & $8.1\pm2.1$ & .. & .. & $-0.475\pm0.065$ & $-0.438\pm0.025$ \\
\enddata
\tablecomments{All times are in seconds since the burst onset time: UT
 09:46:56.1 (GRB\,990123), 01:42:18.7 (GRB\,041219A), 09:29:01.4
 (GRB\,050319), 14:20:06 (GRB\,050401), 01:51:44 (GRB\,050904),
 01:12:15.5 (GRB\,051109A), and 05:59:39 (GRB\,051111). The last two
 events are detailed in
 \S\,\ref{results:main}. Appendix\,\ref{1111tableexplain} gives data
 references and analysis, explaining individual lines event-by-event,
 with data references.  The energy band for the BAT is 15--150 keV,
 and a typical $\gamma$-ray detection energy for a {\it Swift} event
 is $\sim$100 keV. However, lower energy $\gamma$-ray fluxes are used
 when a good measurement is available in order to compare the
 low-energy $\gamma$-ray extension. When a $\gamma$-ray flux density
 is reported with uncertainties larger than $1/3$ of the flux value,
 the detection of high-energy emission in the {\it count rate} is
 greater than 3\,$\sigma$ significant.  The time ranges for
 determining the GRB photon index $\Gamma$ typically extend over the
 entire burst time to get a good measurement, but can be for
 subintervals as indicated in the Appendix. The optical bands are
 ``clear'' (no filter) tied to a filter band, $V$, $R$, $I$ as
 indicated in the subscript. Particularly, the Appendix explains the
 two lines used for GRB\,041219A's first time interval (with two
 estimates of $\beta_{\gamma}$, and the two lines used for GRB\,051111
 (with two optical flux estimates). }
\end{deluxetable}

\end{document}